\RequirePackage{rotating}
\documentclass[useAMS,usenatbib]{mnras}

\usepackage{ulem}
\usepackage{color}
\usepackage{graphics,graphicx}
\usepackage{times} 
\usepackage{amssymb}
\usepackage{amsmath}
\usepackage{natbib}
\usepackage{lscape}
\usepackage{url}
\usepackage{multirow}
\usepackage{ctable}
\usepackage{threeparttable}
\usepackage[labelfont=bf,labelsep=period]{caption}
\newif\ifAMStwofonts
\AMStwofontstrue

\def\xmm{{\it XMM-Newton}}

\def\epicpn{{EPIC-pn}}
\def\epicmos1{{EPIC-MOS1}}
\def\epicmos2{{EPIC-MOS2}}
\def\epicmos{{EPIC-MOS}}

\def\nustar{{\it NuSTAR}}

\def\deg{$^{\circ}$}

\def\kmps{\hbox{$\rm\thinspace km~s^{-1}$}}
\def\pcmsq{\hbox{$\rm cm^{-2}$}}
\def\pcmcub{\hbox{$\rm cm^{-3}$}}
\def\H0{{km~s$^{-1}$~Mpc$^{-1}$}}

\def\kev{\hbox{\rm keV}}

\def\ctps{\hbox{$\rm\thinspace ct~s^{-1}$}}

\def\ergps{\hbox{erg~s$^{-1}$}}
\def\ergcmps{\hbox{\rm erg~cm~s$^{-1}$}}

\def\msun{\hbox{$M_{\odot}$}}

\def\addascaspec{\textsc{addascaspec}}

\def\flx2xsp{\textsc{flx2xsp}}

\def\lcurve{\textsc{lcurve}}

\def\nustardas{\textsc{nustardas}}
\def\nupipeline{\textsc{nupipeline}}
\def\nuproducts{\textsc{nuproducts}}

\def\sas{\textsc{sas}}
\def\xmmselect{\textsc{xmmselect}}

\def\epchain{\textsc{epchain}}
\def\emchain{\textsc{emchain}}

\def\rmfgen{\textsc{rmfgen}}
\def\arfgen{\textsc{arfgen}}
\def\epiclccorr{\textsc{epiclccorr}}

\def\rgsproc{\textsc{rgsproc}}
\def\rgscombine{\textsc{rgscombine}}

\def\chisq{{$\chi^{2}$}}

\def\xspec{\hbox{\small XSPEC}}

\def\xstar{\textsc{xstar}}

\def\tbabs{\textsc{tbabs}}

\def\reflionx{\textsc{reflionx}}
\def\xillver{\textsc{xillver}}
\def\xillvercp{\textsc{xillver\_cp}}

\def\relconv{\textsc{relconv}}
\def\relconvlp{\textsc{relconv\_lp}}
\def\relline{\textsc{relline}}
\def\relxill{\textsc{relxill}}
\def\relxillcp{\textsc{relxillcp}}

\def\relxilllpcp{\textsc{relxilllp\_cp}}

\def\pexrav{\textsc{pexrav}}

\def\borus{\textsc{borus}}

\def\nthcomp{\textsc{nthcomp}}

\def\zashift{\textsc{zashift}}

\def\gabs{\textsc{gabs}}

\def\ovii{\hbox{\rm O\,{\small VII}}}
\def\oviii{\hbox{\rm O\,{\small VIII}}}

\def\ka{$\rm{K}\alpha$}

\def\lya{Ly\,$\alpha$}

\def\eg{{\it e.g.}}

\def\ie{{\it i.e.}}

\def\la{\mathrel{\hbox{\rlap{\hbox{\lower4pt\hbox{$\sim$}}}{\raise2pt\hbox{$<$}}}}}
\def\ga{\mathrel{\hbox{\rlap{\hbox{\lower4pt\hbox{$\sim$}}}{\raise2pt\hbox{$>$}}}}}

\def\d25{D$_{25}$}
\def\nh{{$N_{\rm H}$}}

\def\.25{0.25 keV\thinspace}

\def\kbol210{\rm $\kappa_{2-10}$}

\def\rg{$R_{\rm{G}}$}

\def\rin{$R_{\rm in}$}
\def\rout{$R_{\rm out}$}

\def\Rfrac{$R_{\rm{frac}}$}

\def\pg{PG\,1426+015}
\def\nsims{10,000}
\def\logne{$\log [n_{\rm{e}} / \rm{cm}^{-3}]$}
\def\neresult{$18.0^{+0.1}_{-0.2}$}

\title[The Soft Excess in PG\,1426+015]{The Broadband View of the Bare Seyfert 
PG\,1426+015: Relativistic Reflection, the Soft Excess and the Importance of Oxygen}

\author[D. J. Walton, et al.]
{\parbox{7.in}{D. J. Walton$^{1}$\thanks{E-mail: d.walton4@herts.ac.uk},
A. Madathil-Pottayil$^{1}$,  % comments, DONE
P. Kosec$^{2,3}$, % comments, DONE
J. Jiang$^{4}$, % comments, DONE
J. Garcia$^{5}$, % comments, DONE
A. C. Fabian$^{6}$, % comments, DONE
C. Pinto$^{7}$,  \\ % comments,DONE
D. J. K. Buisson$^{8}$, % comments,DONE
M. L. Parker,$^{9}$ % comments
W. N. Alston$^{1}$, % TBD
C. S. Reynolds$^{10,11}$ % TBD
\\[0.25cm]
\footnotesize
$^{1}$ \it{Centre for Astrophysics Research, University of Hertfordshire, College Lane, Hatfield AL10 9AB, UK} \\
$^{2}$ \it{MIT Kavli Institute for Astrophysics and Space Research, Massachusetts Institute of Technology, Cambridge, MA 02139, USA} \\
$^{3}$ \it{Center for Astrophysics | Harvard \& Smithsonian, Cambridge, MA, USA} \\
$^{4}$ \it{Department of Physics, University of Warwick, Gibbet Hill Road, Coventry CV4 7AL, UK} \\
$^{5}$ \it{NASA Goddard Space Flight Center, Greenbelt, MD 20771, USA} \\
$^{6}$ \it{Institute of Astronomy, University of Cambridge, Madingley Road, Cambridge CB3 0HA, UK} \\
$^{7}$ \it{INAF - IASF Palermo, via Ugo La Malfa 153, 90146 Palermo, Italy} \\
$^{8}$ \it{Independent Researcher} \\
$^{9}$ \it{Optibrium Limited, Cambridge Innovation Park, Cambridge CB25 9GL, UK} \\
$^{10}$ \it{Department of Astronomy, University of Maryland, College Park, MD 20742-2421, USA} \\
$^{11}$ \it{Joint Space-Science Institute, College Park, MD 20742-2421, USA} \\
}}
\date{}

\begin{document}
\pagerange{\pageref{firstpage}--\pageref{lastpage}}
\maketitle
\label{firstpage}

\begin{abstract}
We present results from a deep, coordinated \xmm\ + \nustar\ observation of the
type 1 Seyfert \pg, a source of particular interest as the most massive
reverberation-mapped black hole to date ($\log [M_{\rm{BH}}/M_{\odot}]$ =
$9.01^{+0.11}_{-0.16}$).
The high-resolution RGS data confirm the `bare' nature of the source,
showing no evidence for absorption beyond the Galactic column,  while the broadband
spectrum unambiguously reveals the presence of relativistic reflection from the
innermost accretion disc (in the form of a relativistically broadened iron emission and
associated Compton reflection hump) as well as confirming the presence of the strong
soft excess reported previously. 
We explore whether relativistic reflection can successfully account for the soft excess
along with the higher-energy reflection features, utilizing the two most-commonly used
reflection codes (\reflionx, \xillver). Ultimately we find that both models are able to
successfully reproduce the soft excess, though in the case of the \xillver\ model
this is contingent on reducing the strength of the \oviii\ line included in the model, as
otherwise this feature prevents the model from reproducing the data.
The reflection models that successfully reproduce the broadband data imply a relatively
high density for the accretion disc of \logne\ $\sim 18$, consistent with the loose
anti-correlation seen from other AGN in the \logne\ vs $\log[m_{\rm{BH}} \dot{m}^2]$
plane, as well as a moderate-to-high black hole spin of $a^* \gtrsim 0.7$. This
preliminary spin constraint is strongly dependent on the assumption that the soft
excess is dominated by relativistic reflection.
\end{abstract}

\begin{keywords}
{Galaxies: active -- Black hole physics -- X-rays: individual (\pg)}
\end{keywords}

\section{Introduction}

The reflected emission from the accretion disc provides one of our primary observational
tools for probing the innermost regions around actively accreting black holes, particularly
in the case of the supermassive black holes (SMBHs) powering active galactic nuclei
(where the thermal emission from the disc typically emerges in the UV band, which is
difficult to access owing to interstellar absorption). This arises as a natural
consequence of the disc--corona geometry exhibited by these systems (as some of the
hard X-ray emission from the corona must irradiate the surface of the optically-thick
accretion disc), and carries key information about the innermost radius of the accretion
disc (\eg\ \citealt{kerrconv, relconv}) as well as the geometry of the corona (\eg\
\citealt{Martocchia00, Vaughan04, Wilkins12, Gonzalez17}). The former can in turn provide
information about the spin of the black hole (\citealt{Bardeen72}). This is a property that
is of particular interest for SMBHs, as the growth history of these objects dictates their
distribution in the spin--mass plane (\eg\ \citealt{Sesana14, Bustamante19, Beckmann24}).
A growing number of reflection-based SMBH spin constraints are available in the
literature ($\sim$40--50 to date, see \citealt{Reynolds21rev} and \citealt{Bambi21rev} for
recent reviews). These are mostly limited to relatively low mass black holes ($M_{\rm{BH}}
\lesssim 10^{8}$\,\msun), but spin measurements for the largest black holes are of
particular importance for constraining SMBH growth models (\citealt{Piotrowska24_hexp}).

The reflected emission is dominated by a series of fluorescent emisson lines, among
which the iron line at $\sim$6.4--7.0\,keV (depending on ionisation state) is typically
the strongest, along with a characteristic high energy continuum that peaks at
$\sim$30\,keV and is often referred to as the `Compton hump' (\eg\ \citealt{George91,
reflion, xillver}). While the emission lines are narrow in the rest frame of the disc,
to an external observer they will be subject to the relativistic effects associated
with the orbital motion and the strong gravity close to a black hole, resulting in a
broadened and skewed `diskline' profile (\eg\ \citealt{Fabian89, kdblur}); the detailed
form of this profile is determined by the combined disc--corona geometry. As
such, high quality broadband X-ray spectra are critical to studying the
reflected emission. Indeed, with coordinated \xmm+\nustar\ coverage it is possible to
identify the reflected emission from the disc even in cases with complex (and variable)
absorption (\eg\ \citealt{Risaliti13nat, Walton14, Walton18}; Madathil-Pottayil et al.  2025,
\textit{in preparation}). 

Nevertheless, sources with low obscuration undeniably provide the best view of the
disc reflection features, and are thus particularly valuable for studies of the inner
geometry of AGN and SMBH spin measurements. In addition to the broad iron
emission and the Compton hump, unobscured radio-quiet AGN with appreciable
accretion rates are ubiquitously seen to exhibit a relatively smooth excess of emission
over the extrapolation of the hard X-ray continuum (determined above 2\,keV) down to
lower energies (\eg\ \citealt{Arnaud85,
Gierlinski04, Miniutti09}). This is commonly referred to as the `soft excess'.
The nature of this emission has been hotly debated since its discovery, but one
of the two leading models discussed in the recent literature interprets the soft excess as
another part of the reflection spectrum from the inner disc (\citealt{Crummy06,
Walton13spin, Garcia19}).
Originally this model focused mainly on the forest of emission lines also present
in the reflection spectrum at energies $\lesssim$2\,keV for a broad range of ionisation
states, which can blend together into a relatively smooth emission feature owing to
the relativistic broadening of all of these lines. More recently, though,
reflection models which include the density of the reflector as a free parameter
have become available. This has proven to be a key development, as allowing for disc
densities in the $n \gtrsim 10^{16}$\,\pcmcub\ regime\footnote{Most reflection models
had previously adopted a fixed density of $n = 10^{15}$\,\pcmcub\ (\eg\ \citealt{reflion, 
xillver}).} -- as predicted by disc--corona models for a broad range of mass/accretion
rate parameter space (\citealt{Svensson94}) -- shifts the low-energy free-free
emission in the reflection spectrum up into the observed bandpass, contributing further
soft X-ray flux in addition to the emission lines (\citealt{Garcia16}). The reflection
interpretation has subsequently evolved to include both aspects in its treatment of the
soft excess (\eg\ \citealt{Jiang19agn, Mallick22}).

The other leading model invokes a second, cooler, optically-thick corona to
explain the soft excess (\citealt{Done12, Rozanska15, Petrucci20}). This is typically
interpreted within the framework of a radially-stratified accretion flow. In these models
a slab of optically-thick `warm' Comptonising material shrouds the inner regions of
the blackbody emitting disc, producing the soft excess, in addition to the (even more
centrally located) `hot' corona that produces the hard X-ray emission and the
`standard' disc that emerges at larger radii. As such, this model treats the soft excess
as an additional, distinct emission component contributing to the X-ray spectrum.

As a potential part of the disc reflection spectrum, understanding the origin
of the soft excess has a significant impact on our ability to measure SMBH
spin (and place other geometric constraints on AGN accretion flows).
Temporal lags consistent with the soft excess responding to changes in the higher
energy continuum emission have been seen in many cases (\citealt{FabZog09,
Cackett13, deMarco13, Alston20iras}), and have a natural explanation as reverberation
within the reflection interpretation. An origin that is at least
partially atomic also naturally explains the apparent consistency of the `temperature'
seen for the soft excess over many orders of magnitude in both black hole mass and
accretion rate when treated as a thermal component (\eg\ \citealt{Gierlinski04,
Miniutti09, Bianchi09}). However, with the availability of high S/N broadband spectra
in the \nustar\ era there have been conflicting claims in the literature over whether the
reflection interpretation can reproduce the broadband data, with some analyses
suggesting it can (\eg\ \citealt{Jiang18iras, Xu21, Pottayil24}), and others suggesting it
cannot (even in the era of variable density reflection models; \eg\ \citealt{Porquet18,
Porquet24, Ursini20wc}). In contrast, the `warm' Comptonisation model rarely has
trouble reproducing the spectral data (\eg\ \citealt{Mehdipour15, Petrucci18,
Middei20}), as the soft and hard spectral components are disconnected. As such, the
nature of the soft excess remains unclear, and it may even be possible that both
relativistic reflection and warm Comptonisation are at play (\citealt{rexcor}).

In order to explore these issues further we obtained the first broadband X-ray
observation of \pg\ -- taken with \xmm\ (\citealt{XMM}) and \nustar\ (\citealt{NUSTAR})
in coordination -- and present a detailed analysis of these broadband data in this work.
\pg\ is a nearby ($z=0.08657$), radio-quiet Seyfert-1 AGN which currently stands out
among Seyfert galaxies as having the largest reverberation-mapped mass of any
source included in the AGN Black Hole Mass Database (\citealt{Bentz15}), with
$\log [m_{\rm{BH}}]$ = $9.01^{+0.11}_{-0.16}$ (\citealt{Kaspi00, Peterson04}), where
$m_{\rm{BH}}$ is the black hole mass in solar units (\ie\ $m_{\rm{BH}} = M_{\rm{BH}} / 
M_{\odot}$). The source has a bolometric luminosity of $L_{\rm{bol}} \sim 5 \times 
10^{45}$\,\ergps\ (\citealt{Peterson04})\footnote{Based on a 5100\AA\ luminosity of
$\log [L_{5100} / (\rm{erg}~\rm{s}^{-1})] = 44.72$, and the 5100\AA\ bolometric
correction of 9 adopted in that work.}, corresponding to an Eddington ratio of
$L_{\rm{bol}} / L_{\rm{E}} \sim 0.04$. Although it has received relatively little attention
in the X-ray regime to date, \pg\ appears to be a rare example of a `bare' Seyfert, with
no obvious X-ray absorption beyond the Galactic column. Similar to other bare Seyferts,
\pg\ shows clear evidence for a soft excess below $\sim$2\,keV (\citealt{Porquet04, 
Page04softex}). In terms of the ongoing efforts to populate the mass vs spin plane for
AGN, \pg\ is therefore a source of particular interest, as there are very few high-mass
AGN with robust spin constraints; currently the only robust spin constraints for AGN
likely to be powered by black holes of mass $10^{9}$\,\msun\ or above come from
H1821$+$643 (\citealt{SiskReynes22}) and Q2237$+$305 (\citealt{Reynolds14}).
Furthermore, the unobscured nature of \pg\ provides an ideal opportunity to search for
the key reflection features needed to make such measurements and further explore
whether the reflection interpretation can successfully reproduce the soft excess.

The rest of the paper is structured as follows: Section \ref{sec_red} details our data
reduction and Section \ref{sec_spec} then presents our spectral analysis of the data,
touching on the low energy grating data, the iron \ka\ band and the broadband
continuum in turn. We then discuss our results in Section \ref{sec_dis} and summarise
our conclusions in Section \ref{sec_conc}.

\begin{table}
  \caption{Details of the coordinated \xmm\ and \nustar\ observation of \pg\ considered in
  this work.}
%\vspace{-0.5cm}
\begin{center}
\begin{tabular}{c c c c c}
\hline
\hline
\\[-0.25cm]
Mission & OBSID & Start & Good \\
& & Date & Exposure\tmark[a] \\
\\[-0.3cm]
\hline
\hline
\\[-0.2cm]
\nustar\ & 60501049002 & 2020-01-23 & 105 \\
\\[-0.3cm]
\xmm\ & 0852210101 & 2020-01-24 & 71/99/101 \\
\\[-0.2cm]
\hline
\hline
\\[-0.15cm]
\end{tabular}
\label{tab_obs}
\end{center}
\vspace{-0.2cm}
$^{a}$ Exposures are given in ks, and for \xmm\ are listed for the \epicpn/MOS/RGS 
detectors after background flaring has been excised. Note that all EPIC detectors are
operated in SW mode, resulting in the smaller good exposure for \epicpn\ (which has a
$\sim$70\% livetime in this mode).
%\vspace*{0.3cm}
\end{table}

\section{Observations and Data Reduction}
\label{sec_red}

\nustar\ and \xmm\ performed a coordinated observation of \pg\ in January 2020; a
summary of these observations is given in Table \ref{tab_obs}.

\subsection{\xmm}

The \xmm\ data were reduced using \sas\ v18.0.0. We cleaned the raw observation files
for the EPIC detectors as standard, using \epchain\ for the \epicpn\ detector
(\citealt{XMM_PN}) and \emchain\ for the \epicmos\ detectors (\citealt{XMM_MOS}). All
of the EPIC detectors were operated in Small Window (SW) mode in order to minimise
the risk of pile-up. Source and background products (\ie\ spectra and lightcurves)
were extracted from the resulting cleaned eventfiles with \xmmselect. We use a circular
region of radius 35--40$''$ for the source aperture (the former for \epicpn, the latter
for \epicmos), and a larger region of blank sky to estimate the background. For the
\epicpn\ detector the background region is placed on the same chip as \pg, but the use
of small window mode means this is not possible for the \epicmos\ detectors, and so
these regions are placed on adjacent chips. As recommended,  periods of particularly
high background were excluded from our analysis; for these data, this just removes a
few ks of exposure right at the end of the observation. We only include single and
double patterned events for \epicpn\ ({\small PATTERN}\,$\leq$\,4) and single to
quadruple patterned events for \epicmos\ ({\small PATTERN}\,$\leq$\,12) in our analysis,
and the appropriate instrumental response files were generated for each detector using
\rmfgen\ for the redistribution matrices and \arfgen\ for the ancillary response files.
Although we performed the reduction for each of the two \epicmos\ units, the spectral
data from these detectors were then combined into a single \epicmos\ spectrum using
\addascaspec. Source lightcurves were corrected for PSF losses based on the size of
the source aperture using \epiclccorr. The total count rates
($\sim$5--6.5\,\ctps\ for \epicpn\ and $\sim$1.2--1.7\,\ctps\ for each \epicmos\
detector) were sufficiently low that pile-up is of no concern owing to the use of SW
mode.

\begin{figure}
\begin{center}
\hspace*{-0.35cm}
\rotatebox{0}{
{\includegraphics[width=235pt]{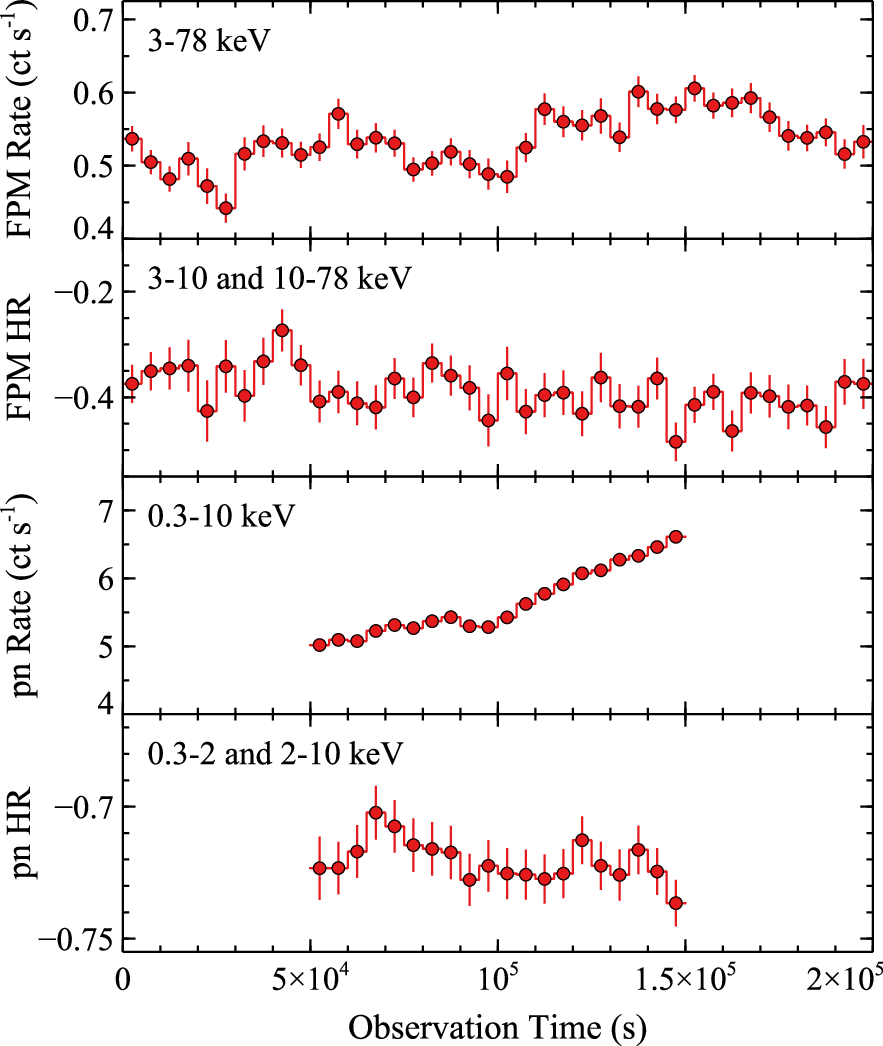}}
}
\end{center}
\vspace*{-0.3cm}
\caption{Lightcurves and hardness ratios (HRs) for the \nustar\ (top two panels) and
\xmm\ (bottom two panels) observations considered here. The HRs shown here are
defined as $(H-S)/(H+S)$, where $H$ and $S$ are the count rates in the harder and
softer of the indicated bands, respectively, and the bands used are defined in the
observed frame. For \nustar\ we show the combined data for FPMA+FPMB, and for
\xmm\ we show the data for the EPIC-pn detector.
}
\label{fig_lc}
\end{figure}

The data from the Reflection Grating Spectrometer (RGS; \citealt{XMM_RGS}) were
reduced using \rgsproc, which both cleans the raw event files and extracts the
spectral products and their associated instrumental response files. Both of the RGS
units were operated in standard Spectroscopy mode, and we used both the standard
source and background regions. As with the EPIC data, periods of particularly high
background were excluded. The net source count rates were $\sim$0.15\,\ctps\ for
each RGS detector, and the data from RGS1 and RGS2 were merged into a single
RGS spectrum using the \rgscombine\ routine (after confirming there were no
notable differences between them over their common energy coverage).

\subsection{\nustar}

The \nustar\ data were reduced following standard procedures with the \nustar\ Data
Analysis Software (\nustardas) v1.9.2, and \nustar\ calibration database v20200726. 
The unfiltered event files for both focal plane modules (FPMs) A and B were cleaned
with \nupipeline, using the standard depth correction to reduce the internal high-energy
background. Passages through the South Atlantic Anomaly were excluded using the
minimal, default settings. Source products and their associated instrumental response
files were then extracted for each module using circular regions of radius 80$''$ with
\nuproducts. As with the \xmm\ data the background was estimated from larger regions
of blank sky, which were placed on the same chip as \pg. We extract both the standard
`science' data (mode 1) and the `spacecraft science' data (mode 6; see
\citealt{Walton16cyg}) in order to maximise the signal-to-noise (S/N); the mode 6 data
provide $\sim$10\% of the total \nustar\ exposure quoted in Table \ref{tab_obs} in this
case. While the optics temperatures indicate this observation could potentially
be affected by the tear in the thermal blanket that was recently identified
(\citealt{NuSTARmli}), which can degrade the consistency of the two FPMs at the
lowest energies covered by \nustar, we see no evidence for disagreement between
FPMA and FPMB in this case and so we use the \nustar\ data over the full 3--78\,keV
bandpass.  Finally, we merged the FPMA and FPMB data into a combined FPM spectrum
using \addascaspec, and similarly summed the FPMA and FPMB lightcurves into
a combined FPM lightcurve using \lcurve.

\begin{figure*}
\begin{center}
\hspace*{-0.4cm}
\rotatebox{0}{
{\includegraphics[width=235pt]{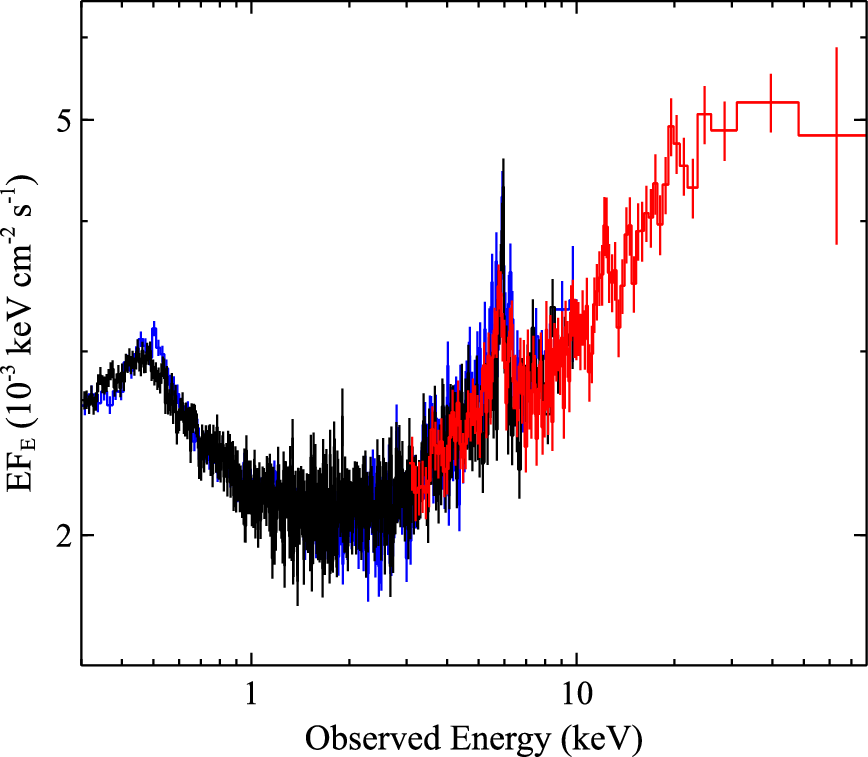}}
}
\hspace*{0.6cm}
\rotatebox{0}{
{\includegraphics[width=235pt]{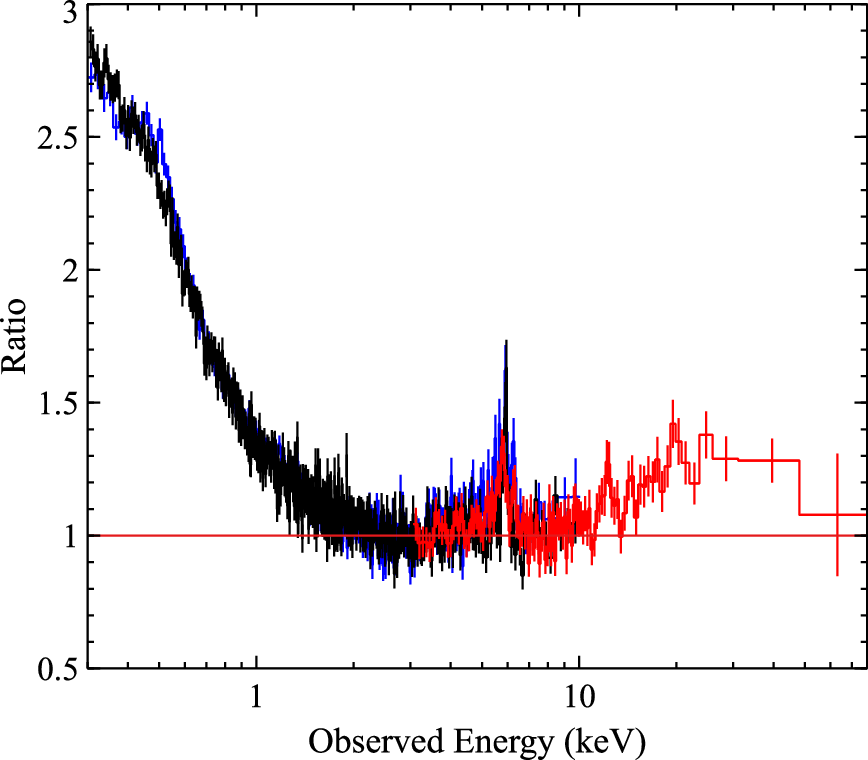}}
}
\end{center}
\vspace*{-0.3cm}
\caption{
Left panel: the broadband \xmm+\nustar\ spectrum of \pg\ (after being unfolded through
a model that is constant in $EF_{\rm{E}}$). The \xmm\ data are shown in black and blue
(\epicpn\ and \epicmos, respectively) and the \nustar\ data are shown in red (combined
FPMA and FPMB data). Right panel: residuals to a simple powerlaw continuum
($\Gamma = 1.76$), modified by Galactic absorption and applied to the broadband data
over the 2--4, 7--10 and 50--78\,\kev\ energy ranges (bands that should be
dominated by the primary AGN continuum). A strong and smooth soft excess can be
seen below $\sim$2\,keV, as is also seen in the archival \xmm\ data (\citealt{Porquet04, 
Page04softex}) and other unobscured AGN (\eg\ \citealt{Walton13spin}). In addition, iron
emission can clearly be seen, along with a hard excess above $\sim$10\,keV indicating
the presence of a reflection continuum. Note that the data in both panels have been
rebinned for visual clarity.}
\label{fig_spec}
\end{figure*}

\section{X-ray Spectroscopy}
\label{sec_spec}

Limited variability is seen by either \xmm\ or \nustar\ during this new observation (see
Figure \ref{fig_lc}); the data show a systematic rising trend, with an overall increase in
flux of $\sim$20--30\% seen in both cases, but there is no evidence for strong
associated spectral variability based on lightcurves extracted in the 0.3--2.0 and
2.0--10.0\,keV bands for \xmm, and in the 3--10 and 10--78\,keV bands for \nustar.
We therefore focus our analysis on the time-averaged data. All of the EPIC and
\nustar\ datasets are binned to a minimum signal-to-noise (S/N) of 5. Our spectral
analysis is performed with \xspec\ (\citealt{xspec}), and unless stated otherwise we fit
the data by minimising the \chisq\ statistic. All of our spectral models include
absorption from the Galactic column in the direction of \pg\ ($N_{\rm{H,Gal}} = 2.6
\times 10^{20}$\,\pcmsq; \citealt{NH2016}). This is modelled with the \tbabs\ 
absorption code (\citealt{tbabs}), and we assume the absorption cross-sections of
\cite{Verner96} and the solar abundances of \cite{Grevesse98}, since these
abundances are implicitly assumed in many of the spectral models we use throughout
this work. Additionally, for the fits that involve both the \xmm\ and \nustar\ data
(Sections \ref{sec_fek}--\ref{sec_broad}) we allow for multiplicative normalisation
constants to vary for the \epicmos\ and FPM data (with the constant for the \epicpn\
data fixed to unity) to account for any cross-calibration issues, as well as the slightly
different durations of the exposures, fixing the EPIC-pn detector at unity.  These are
always found to be within $\sim$5\% of unity.

\begin{figure}
\begin{center}
\hspace*{-0.35cm}
\rotatebox{0}{
{\includegraphics[width=235pt]{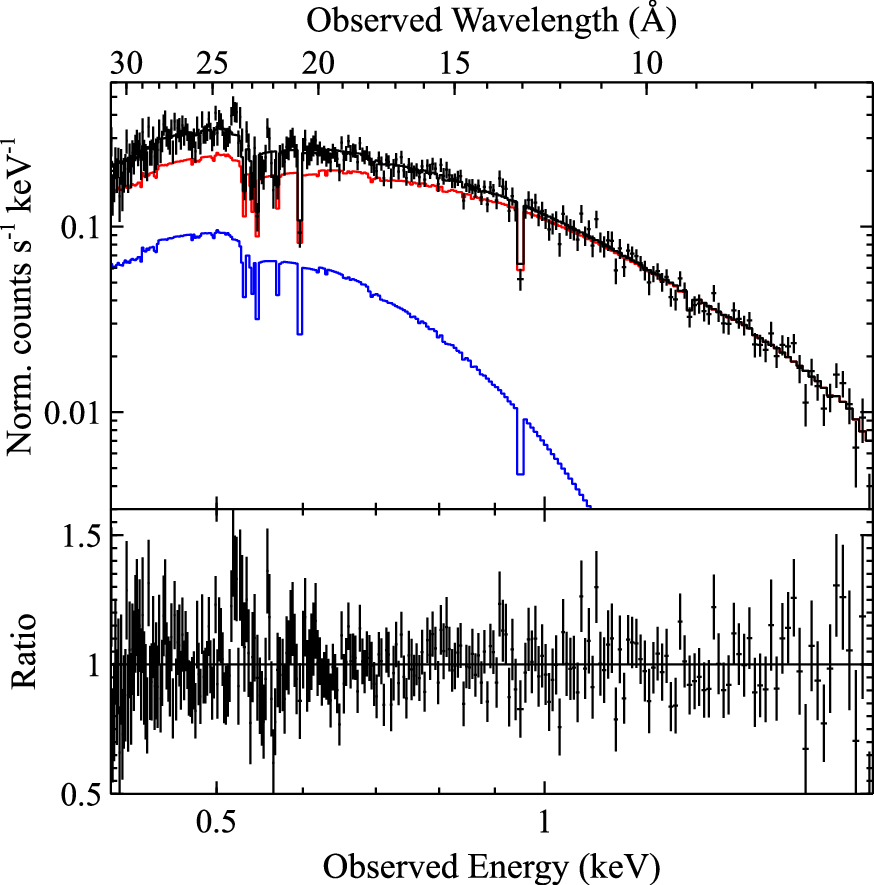}}
}
\end{center}
\vspace*{-0.3cm}
\caption{The fit to the RGS data with a simple phenomenological continuum model
consisting of a blackbody and a powerlaw, modified by only Galactic absorption.
The top panel shows the fit to the raw data, with the black, blue and red lines
showing the total model, the blackbody and the powerlaw, respectively. The bottom
panel shows the data/model ratio for this model. The data have been re-binned for
visual purposes.
}
\label{fig_rgs_fit}
\end{figure}

\begin{figure*}
\begin{center}
\hspace*{-0.4cm}
\rotatebox{0}{
{\includegraphics[width=470pt]{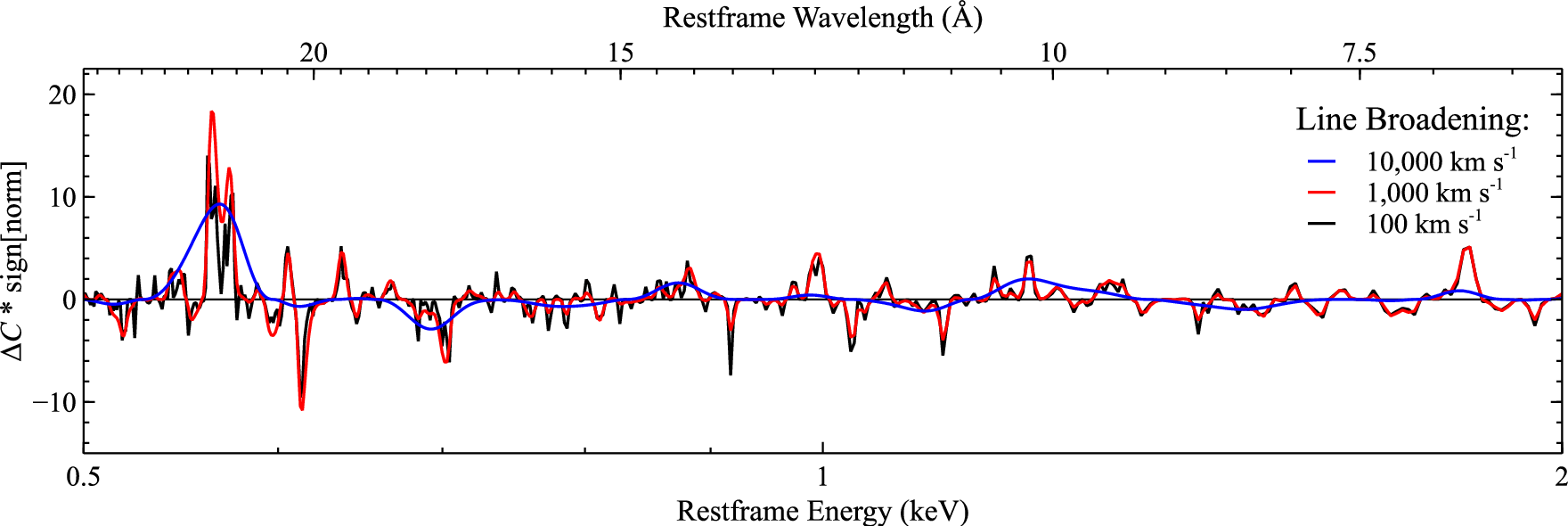}}
}
\end{center}
\vspace*{-0.3cm}
\caption{The results from a set of Gaussian line scans applied to the RGS data,
taking the continuum fit shown in Fig. \ref{fig_rgs_fit} as our baseline model, and
assuming a few different line widths for illustration. Line energies are scanned
across the 6.2--24.8\,\AA\ bandpass in steps of 0.04\,\AA\ (roughly the spectral
resolution of the 1$^{\rm{st}}$ order RGS data considered here), with the 
improvement ($\Delta C$) noted for each step. Positive values of $\Delta C$
indicate emission features, and negative values indicate absorption. The only
feature of any note is \ovii\ emission at 0.57\,keV.
}
\label{fig_rgs_scan}
\end{figure*}

The broadband spectrum is shown in Figure \ref{fig_spec} (left panel). In order to
highlight the main features present in the spectrum, we also show the data relative to
a simple powerlaw continuum, fit to the 2--4, 7--10 and 50--78\,keV energy bands
(rest frame), as has become standard since these energy ranges should be
dominated by the primary AGN continuum. In general the spectrum is moderately hard,
rising in EF$_{\rm{E}}$ space above $\sim$2\,keV; the best-fit photon index in the
powerlaw model is $\Gamma = 1.76$, fairly typical for AGN at moderate accretion
rates (\eg\ \citealt{Brightman13, Ricci17}). At lower energies a strong soft excess is
present, as seen in the short archival \xmm\ observation (\citealt{Porquet04,
Page04softex}), and there is no indication for any additional absorption associated
with \pg\ in the EPIC data. Iron emission is also clearly seen at $\sim$6.4 keV in the
rest frame of \pg, and there is a hard excess seen above 10\,keV clearly
demonstrating the presence of a high-energy reflection continuum (\ie the `Compton 
hump').

\subsection{The RGS Data}
\label{sec_rgs}

Given the apparent lack of absorption in the EPIC data, before embarking on a detailed
analysis of the broadband spectrum we first inspect the RGS data in order to robustly
confirm the nature of \pg\ as a `bare' Seyfert. In order to do so, we fit the RGS data
with a phenomenological continuum model consisting of a blackbody and a powerlaw,
both of which are modified by Galactic absorption. \pg\ is well detected in the RGS data
over the 0.4--2.0\,keV band in the observed frame. Owing to the lower count rates and
higher resolution in comparison to the EPIC data, we bin the RGS data to a minimum
of 1 count per bin to facilitate the use of the Cash statistic\footnote{This is functionally
similar to the \chisq\ statistic, but is defined specifically for use with measurement
errors that are dominated by Poisson counting statistics (instead of assuming Gaussian
errors),  properly accounting for the asymmetric errors associated with the low-count
regime relevant for the RGS data analysed here.} (\citealt{cstat}) within \xspec\ while
retaining the maximum spectral resolution. Our simple continuum describes the RGS
data well (see Figure \ref{fig_rgs_fit}), with $C = 2518$ for 2433 degrees of freedom
(DoF). We find a blackbody temperature of $kT = 0.11 \pm 0.01$\,keV, broadly similar to
the temperatures seen for other AGN when the soft excess is modeled as a blackbody
(\eg\ \citealt{Gierlinski04, Crummy06, Miniutti09}). Even with the blackbody included, we
find a softer photon index of $\Gamma = 2.20 \pm 0.15$ in the RGS band than seen
above 2\,keV in the EPIC data.

We then search for evidence for any atomic features (either in absorption or emission)
by scanning a Gaussian feature across the majority of the observed bandpass.
Formally, we step the line energy across the 6.2--24.8\,\AA\ wavelength range
($\sim$0.5--2.0\,keV) in the rest-frame of \pg\ in steps of 0.04\,\AA, broadly equivalent
to the spectral resolution of the RGS data considered here. We explored a range of
values for the line broadening ranging from 100--10,000\,\kmps, similar to \eg\
\cite{Kosec20}, but only show a few examples in Figure \ref{fig_rgs_scan} for
illustration. At each step we fit for the line normalisation, which can be either positive
(emission) or negative (absorption). The only feature of any note is an emission
feature at a rest-frame energy of $\sim$0.57\,keV, which can be identified as the \ovii\
\lya\ triplet. Modelling this as a single line with a full set of free parameters, the fit is
improved by $\Delta C = 20$ for three additional free parameters, and we find a line
energy, width and equivalent width of $E = 569^{+3}_{-6}$\,eV, $\sigma =
5.4^{+2.4}_{-1.8}$\,eV (corresponding to a line broadening of $\sim$3000\,\kmps) and
$EW = 4.7 \pm 2.1$\,eV. Allowing instead for a full treatment of the fact that this feature
is a triplet -- fixing the line energies at 561, 569 and 574\,eV for the $f$, $i$ and $r$
transitions, respectively, and assuming a common line broadening -- does not improve
the fit further, and results in strong degeneracies between the different line
normalisations.

\begin{figure}
\begin{center}
\hspace*{-0.35cm}
\rotatebox{0}{
{\includegraphics[width=235pt]{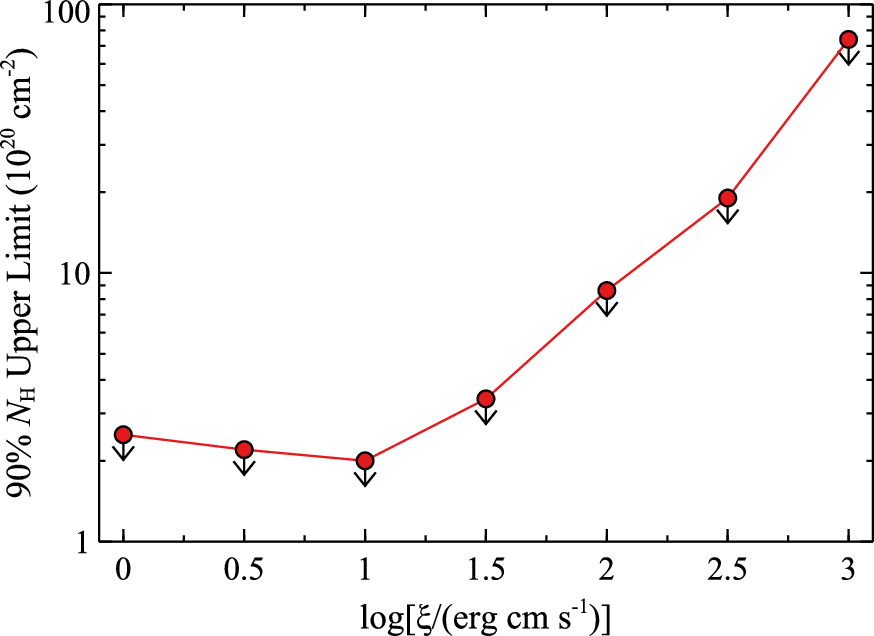}}
}
\end{center}
\vspace*{-0.3cm}
\caption{Upper limits from the RGS data on the column density of any ionised
absorption towards \pg\ as a function of ionisation parameter. These have been
calculated using an \xstar\ absorption model that assumes a $\Gamma = 2$
powerlaw ionising continuum, a velocity  broadening of 100\,\kmps\ and solar
abundances.
}
\label{fig_rgs_limits}
\end{figure}

Critically, though, there is no obvious indication from any of the line scans performed
for any absorption beyond the Galactic column in the RGS data. We also performed
some simple tests using full absorption models, rather than individual line scans.
First, we allowed for the presence of a second neutral absorber at the redshift of \pg,
assuming solar abundances. This offered no improvement to the fit, and we found
an upper limit on the column density of $N_{\rm{H}} < 8 \times 10^{19}$\,\pcmsq. We
also allowed for an ionised absorber using \xstar\ (\citealt{xstar}), relying on a
pre-calculated grid of absorption models from \cite{Walton20iras} computed
assuming a powerlaw ionising continuum with $\Gamma = 2$ and a velocity
broadening of 100\,\kmps\ (fairly typical of `warm' absorbers seen in other AGN, \eg\
\citealt{Laha14}). The absorber is allowed to be blueshifted relative to \pg, and we
again assume solar abundances.\footnote{\xstar\ uses the solar abundance set of
\cite{Grevesse98}.} This also only offered marginal improvements in the fit, and
across a range of ionisation parameters spanning
$\log[\xi/(\rm{erg}~\rm{cm}~\rm{s}^{-1})]$ = 0.0--3.0 (also typical for warm
absorbers) we always find upper limits on the column density of $N_{\rm{H}} \lesssim
7 \times 10^{21}$\,\pcmsq; the upper limits as a function of $\xi$ are shown in Fig.
\ref{fig_rgs_limits} (note that the \xstar\ grid itself has a lower bound on the column
density of $10^{20}$\,\pcmsq). Here, $\xi$ has its usual definition: $\xi =
L_{\rm{ion}}/nR^{2}$, where $L_{\rm{ion}}$ is the ionising luminosity, $n$ is the density
of the absorber and $R$ is the distance to the ionising source. Taken altogether, we
conclude that these tests confirm \pg\ as a rare example of an essentially completely
bare AGN. As such, we do not consider the RGS data any further here.

\subsection{The Iron K Band}
\label{sec_fek}

\begin{figure}
\begin{center}
\hspace*{-0.35cm}
\rotatebox{0}{
{\includegraphics[width=235pt]{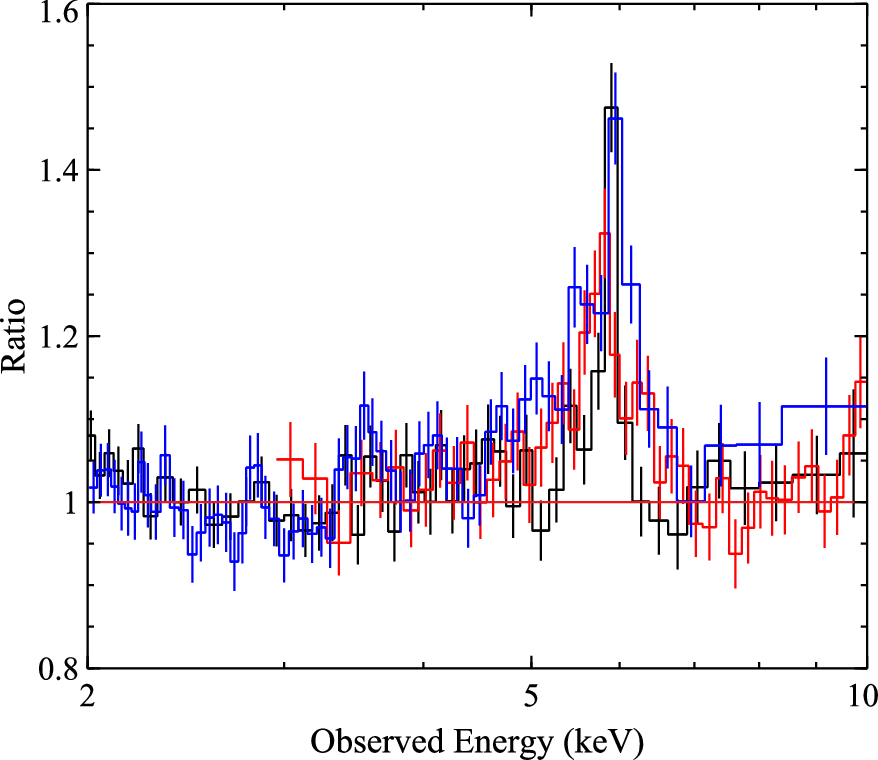}}
}
\end{center}
\vspace*{-0.3cm}
\caption{A zoom-in on the 2--10\,keV band for the data/model ratio to the simple
powerlaw continuum also shown in Figure \ref{fig_spec}, highlighting the iron K
residuals. The colours have the same meaning as Figure \ref{fig_spec}, and the data
have again been rebinned for visual clarity. Overall, the data appear to show evidence
for both a narrow core to the iron emission, and a broad, relativistic component
underlying this. However, regarding the latter, there is some disagreement between
the different detectors; notably the broad component is not as obvious in the
\epicpn\ data. Nevertheless, statistical analysis indicates this broad component is
significantly detected when considering all of the available data (see Section
\ref{sec_fek}).
}
\label{fig_FeK}
\end{figure}

We also present an inspection of the iron K band to test for the presence of relativistic
reflection. In Figure \ref{fig_FeK} we show a zoom-in on the 2--10\,keV for the powerlaw
fit to the high-energy data described above (\ie\ $\Gamma = 1.76$). The combined data
indicate the presence of a broad, relativistic component to the iron line, in addition to a
narrow core.

In order to assess the presence of this broad component we test two models for the
2--10\,keV band, fit to the combined \xmm+\nustar\ data simultaneously: first, a
powerlaw continuum with a narrow Gaussian emission line at 6.4\,keV (in the rest-frame
of \pg), and second the same model with a relativistic line added. In the latter case, we
use the \relline\ model (\citealt{relconv}). For the \relline\ component, we allow the line
energy, the black hole spin and the disc inclination to vary (although the line energy is
restricted to 6.4--6.97\,keV, corresponding to neutral and hydrogen-like iron), and we
assume a powerlaw emissivity profile, with $\epsilon(r) \propto r^{-q}$, where the
emissivity index $q$ is also a free parameter. Although the simpler model (powerlaw
with a narrow Gaussian) fits the data fairly well, with \chisq\ = 1527 for 1377 degrees of
freedom (DoF), the inclusion of the \relline\ component further improves the fit by
$\Delta\chi^{2} = 69$ for 5 extra free parameters. This suggests a very significant
statistical improvement. The narrow core is consistent with neutral iron, with an
energy of $E = 6.415 \pm 0.015$\,keV, and has an equivalent width of $EW = 66 \pm
16$\,eV, typical of other unobscured, radio-quiet AGN (\eg\ \citealt{Bianchi07}). The
broad, relativistic component has $EW = 200 \pm 60$\,eV, similar to that expected for
standard illumination of an accretion disc (\eg\ \citealt{George91}).

In order to assess the statistical significance of the broad iron component we
performed a series of Monte Carlo simulations, largely following a similar approach to
\cite{Walton15lqso}. In short, using the {\small{fakeit}} command in \xspec, we
simulated \nsims\ sets of \xmm\ and \nustar\ spectra with the same characteristics
(exposures, backgrounds, binning criterion) as the real data, using the best-fit model
without the \relline\ component (\ie\ just a powerlaw continuum with Galactic
absorption and a narrow iron emission line) as our input. We then model these
datasets over the 2--10\,keV band before and after including an additional \relline\
component, determining the improvement in \chisq\ in order to assess  the probability
of obtaining the observed $\Delta\chi^{2}$ by chance. In these analyses, the \relline\
component is treated in exactly the same way as in our analysis of the real data. Of
the \nsims\ datasets simulated, none show a chance improvement equivalent to or
greater than that observed. In fact, the maximum improvement we obtain by chance
in these simulations is $\Delta\chi^{2} = 24.3$, significantly below that seen in the real
data. This implies the formal detection significance of the broad iron component
comfortably exceeds the 4$\sigma$ level, strongly supporting the visual impression
that a relativistic component to the iron line is present in \pg.

\subsection{Hard X-ray Modelling}
\label{sec_hard}

Having established the presence of relativistic disc reflection in the spectrum of \pg\
(in addition to the relativistic iron line discussed in the previous section, we also note
again the presence of a clear Compton hump in the broadband residuals shown in Fig.
\ref{fig_spec}), we now bring in the higher energy \nustar\ data and model the hard
X-ray data ($>$2\,keV in the rest-frame of \pg) with physical reflection models. 
In order to describe the relativistic disc reflection we make use of the two primary
reflection models used in the literature: the \xillver\ model (\citealt{xillver}) and the
\reflionx\ model (\citealt{reflion}). In both cases, the relativistic effects relevant to
the innermost accretion disc are applied by convolving these reflection models with
the \relconv\ model (\citealt{relconv}).

\begin{table*}
  \caption{Parameter constraints for the disc reflection model fit to
  the hard X-ray data available for \pg.}
\begin{center}
%\vspace{-0.2cm}
\begin{tabular}{c c c c c c c}
\hline
\hline
\\[-0.2cm]
Model Component & \multicolumn{2}{c}{Parameter} & \relxilllpcp\ & \relconvlp$\otimes$\reflionx \\
\\[-0.2cm]
\hline
\\[-0.2cm]
Primary Continuum & $\Gamma$ & & $1.92^{+0.04}_{-0.05}$ & $1.91^{+0.04}_{-0.16}$ \\ 
\\[-0.3cm]
(\nthcomp) & $kT_{\rm{e}}$\tmark[a] & [keV] & $>120$ & $>60$ \\
\\[-0.3cm]
& Norm\tmark[b] & [$10^{-3}$] & -- & $2.1^{+0.1}_{-1.1}$ \\
\\[-0.05cm]
Disc Reflection & $a^*$ & & $> -0.17$ & $>0.07$ \\
\\[-0.3cm]
(\relxilllpcp\ or & $i$ & [\deg] & $33 \pm 3$ & $<39$ \\
\\[-0.3cm]
\relconvlp$\otimes$\reflionx) & $h$ & [\rg] & $<7.3$ & $<4.8$ \\
\\[-0.3cm]
& $A_{\rm{Fe}}$ & [solar] & $1.7^{+0.6}_{-0.5}$ & $1.7^{+0.3}_{-0.9}$ \\
\\[-0.3cm]
& $\log\xi$ & $\log$[\ergcmps] & $1.8^{+0.3}_{-0.2}$ & $1.7^{+1.5}_{-0.3}$ \\
\\[-0.3cm]
& \Rfrac\tmark[c] & & $1.3 \pm 0.4$ & $0.8^{+0.6}_{-0.2}$ \\
\\[-0.3cm]
& Norm & & $1.2^{+2.8}_{-0.6} \times 10^{-4}$ & $1.4^{+0.4}_{-1.1}$ \\
\\[-0.05cm]
Distant Reflection & $\log$\,\nh\ & log[\pcmsq] & $23.2^{+0.7}_{-0.3}$ & $23.1^{+1.4}_{-0.1}$ \\
\\[-0.3cm]
(\borus) & $C_{\rm{tor}}$ & [4$\pi$] & $0.82^{+0.04}_{-0.55}$ & $0.32^{+0.58}_{-0.07}$ \\
\\[-0.2cm]
\hline
\\[-0.25cm]
\chisq/DoF & & & 1722/1645 & 1714/1645 \\
\\[-0.25cm]
\hline
\hline
\end{tabular}
\label{tab_hard_param}
\end{center}
\flushleft
$^a$ $kT_{\rm{e}}$ is quoted in the rest-frame of the X-ray source (\ie prior to any
gravitational redshift), based on the best-fit lamppost geometry. \\
$^b$ The primary \nthcomp\ continuum is incorporated within the \relxilllpcp\ model,
and so does not have a separate normalisation. \\
$^c$ \Rfrac\ is not formally a free parameter for the \relconvlp$\otimes$\reflionx\
model, but is calculated following the method outlined in \cite{Walton13spin}, i.e. via
a comparison with the \pexrav\ model (\citealt{pexrav}). \\
%\vspace{0.2cm}
\end{table*}

The self-consistent combination of \relconv\ and \xillver\ is implemented within the
\relxill\ family of models (\citealt{relxill}; we make use of v2.3 here. Specifically, we use
the \relxilllpcp\ model, which adopts the \nthcomp\ thermal Comptonization model
(\citealt{nthcomp1, nthcomp2}) as the ionising continuum, and assumes a lamppost
geometry for the disc--corona system. This geometry is somewhat simplistic, but its
use nevertheless has some advantages in that it allows us to exclude regions of
parameter space that would typically be considered non-physical (\eg\ a very steep
radial emissivity profile and a non-rotating black hole) and it also allows for a physical
interpretation of the reflection fraction, \Rfrac, which determines the relative
normalisations of the ionising continuum and the reflected emission (see
\citealt{relxill_norm}). We note explicitly that the \relxilllpcp\ model includes both the
contributions from the ionising continuum and the reflection from the accretion disc.
The \nthcomp\ continuum is characterised by the photon index and the electron
temperature of the Comptonizing region. The other key parameters associated with
the reflection (in addition to \Rfrac) included in the model are the spin of the black hole
($a^*$), the inclination, density, inner and outer radii, iron abundance\footnote{The
rest of the elements are assumed to have solar abundances.} and ionisation state of
the disc ($i$, $n$, \rin, \rout, $A_{\rm{Fe}}$, $\xi$), and the height of the X-ray
source above the spin axis ($h$). In this analysis we assume that the accretion disc
extends down to the innermost stable circular orbit (ISCO), and set the outer radius of
the disc to the maximum value allowed by the model (1000\,\rg, where \rg\ $= GM/c^2$
is the gravitational radius). We also allow for the treatment of returning radiation
included in the latest versions of the model (see \citealt{relxill_rtnrad}), and for
simplicity in these hard X-ray fits we also assume a `standard' disc density of $n = 
10^{15}$\,\pcmcub.

The model utilizing \reflionx\ for the disc reflection has most of the same parameters
as described above. We again make use of a version of \reflionx\ that assumes an
\nthcomp\ ionising continuum, although here the primary continuum needs to be
treated separately and so a further \nthcomp\ component is included in the model.
The parameters for the primary continuum are linked to those of the \reflionx\ model;
in the case of the electron temperature, we do so after accounting for the gravitational
redshift implied by the combination of $a^*$ and $h$, since with the model setup
utilized here the temperature parameter for the \reflionx\ component is evaluated in
the rest-frame of the corona while the temperature parameter for the \nthcomp\
component is evaluated in the frame of an external observer. We also retain the main
assumptions outlined above, and again assume a lamppost geometry for the corona
(through the use of the \relconvlp\ convolution model specifically), although since the
reflection and relativistic blurring components are distinct model components here
there is no treatment of returning radiation considered.

As the X-ray spectrum of \pg\ also shows a clear narrow core to the iron emission (Fig.
\ref{fig_FeK}), we also allow for the presence of more distant reprocessing (in addtion
to the disc reflection) by more distant material. Here, we use the \borus\ model
(\citealt{borus}), a physically self-consistent model for reprocessing in a distant
torus-like structure (formally the geometry is spherical with conical polar cutouts). The
version we use also adopts an \nthcomp\ model for the ionising continuum, and we link
these parameters to those of the disc reflection components in the two models we
consider here (again, in the case of the electron temperature we do so after accounting
for the gravitational redshift implied by $a^*$ and $h$). The other key parameters are
the column density through the torus (\nh; note that \borus\ allows for both the
Compton-thin and Compton-thick regimes), its covering factor ($C_{\rm{tor}} = 
\Omega_{\rm{tor}} / 4\pi$, where $\Omega_{\rm{tor}}$ is the solid angle subtended by
the torus, such that a spherical obscurer would have $C_{\rm{tor}} = 1$) and iron
abundance ($A_{\rm{Fe}}$), and the angle it is viewed at (which is also linked to the
inclination in the disc reflection models). Given the unobscured nature of \pg, we set a
requirement that we cannot be viewing the source through the torus (i.e. $C_{\rm{tor}}
< \cos i$). The iron abundance is also linked to the disc reflection model, after
correcting for the different solar abundances assumed in these models.\footnote{The
\relxill\ family of models also adopts the solar abundances of \cite{Grevesse98}, the
\reflionx\ model adopts the solar abundances of \cite{Morrison83}, and the \borus\
model adopts the solar abundances of \cite{Anders89}. The iron abundance (relative
to hydrogen) is a factor of 0.68 and 0.71 lower in the first two abundance sets,
respectively, than in the latter, so these are the correction factors we apply.}

\begin{figure}
\begin{center}
\hspace*{-0.35cm}
\rotatebox{0}{
{\includegraphics[width=235pt]{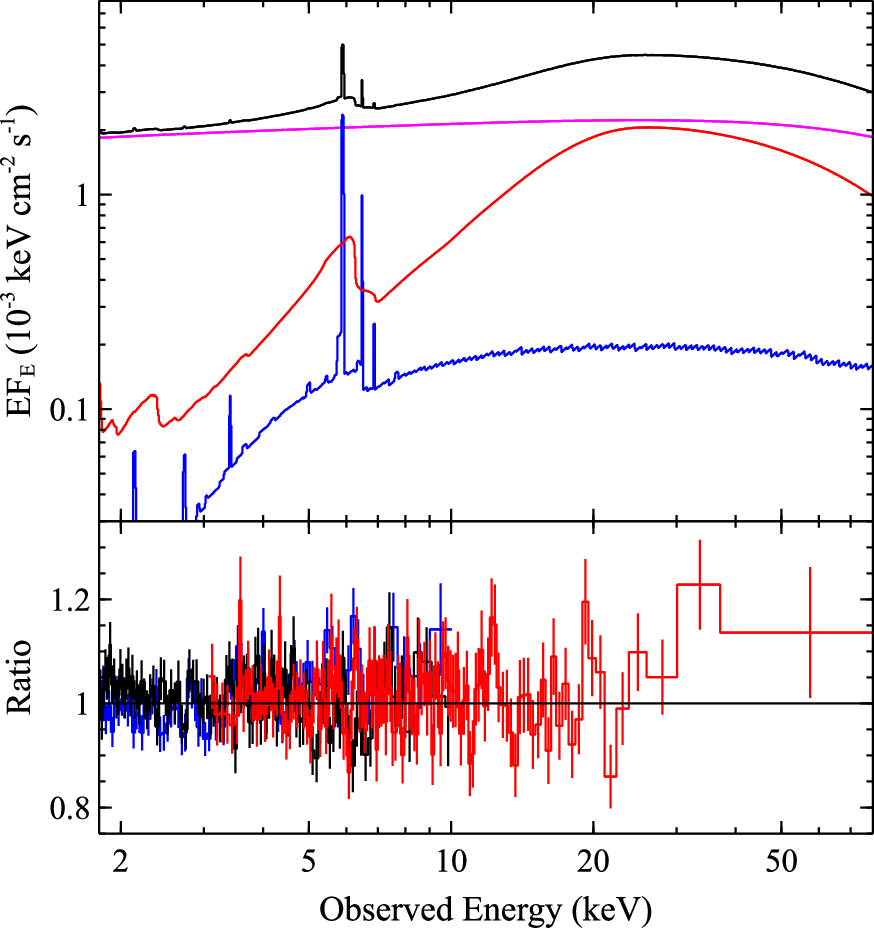}}
}
\end{center}
\vspace*{-0.3cm}
\caption{\textit{top panel:} the best-fit \relxilllpcp+\borus\ model for the hard X-ray
data of \pg\ ($>$2\,keV in its rest frame). The black curve shows the total model, while
the magenta, red and blue curves show the relative contributions of the primary
continuum, the relativistic disc reflection and the distant reprocessing, respectively.
\textit{bottom panel:} the data/model ratios for this fit. Here the colours have the same
meaning as Fig. \ref{fig_spec}, and the data have been further rebinned for visual
clarity.
}
\label{fig_hard_fit}
\end{figure}

\begin{figure*}
\begin{center}
\hspace*{-0.35cm}
\rotatebox{0}{
{\includegraphics[width=470pt]{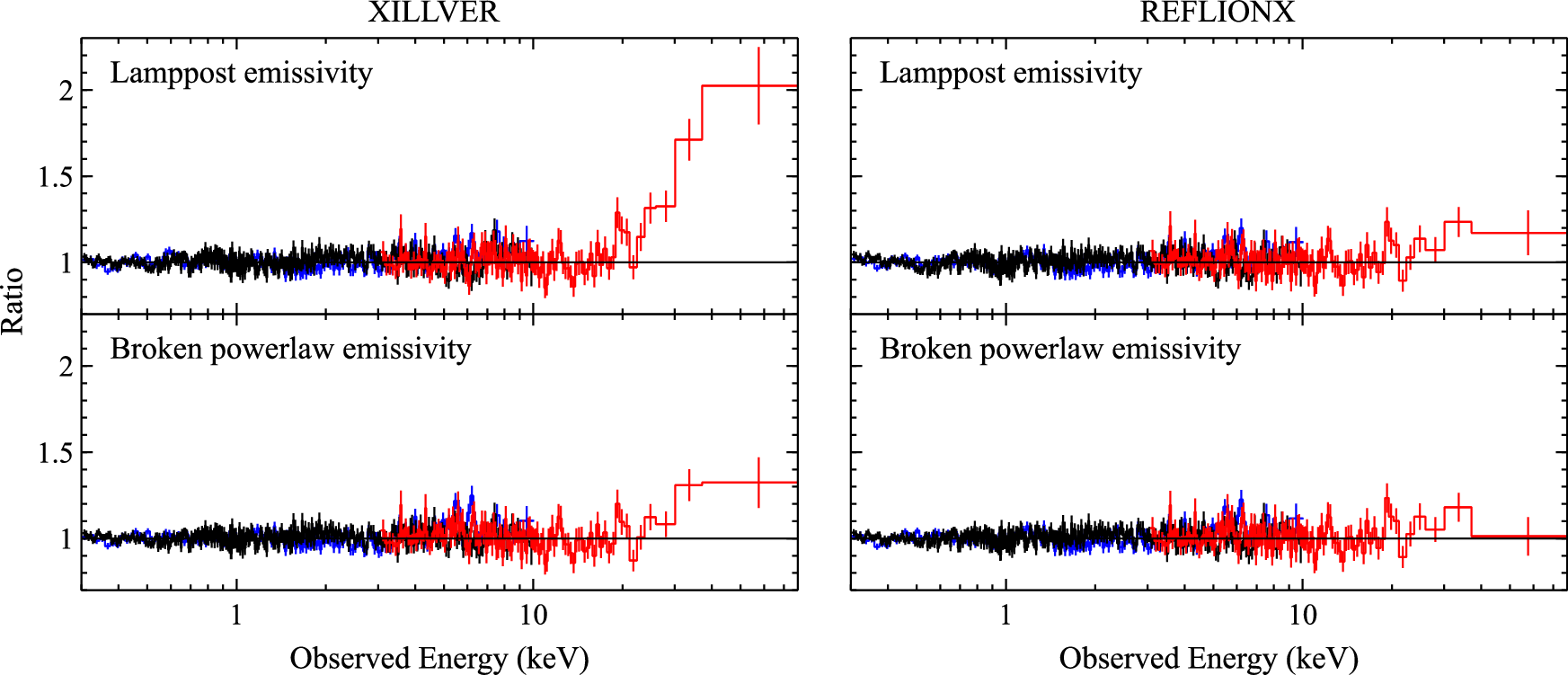}}
}
\end{center}
\vspace*{-0.3cm}
\caption{Data/model ratios for the set of `basic' disc reflection models fit to the full
broadband dataset. The left panels show the models based on the \xillver\ reflection
model (specifically using the \relxill\ family of models) and the right panels show the
models based on the \reflionx\ models. The top panels show the fits assuming a
lamppost emissivity, and the bottom panels show the fits assuming a broken
powerlaw emissivity which is agnostic about the emission source geometry. The
colours again have the same meaning as Fig. \ref{fig_spec}, and the data have been
further rebinned for visual clarity.
}
\label{fig_broad_ratio}
\end{figure*}

Use of a physical torus model for the distant reflector allows the overall spectral
model to incorporate the subtle changes in the strength and shape of its Compton
hump as $C_{\rm{tor}}$ varies, and in turn help constrain this parameter. To further
help constrain $C_{\rm{tor}}$ the \borus\ normalisation should in principle be linked
to that of the ionising continuum (as otherwise these two parameters are often highly
degenerate, since they both control the absolute flux of the distant reflection). For the
\reflionx-based model we therefore link the \borus\ normalisation to that of the
\nthcomp\ component (after accounting for the fact that the \nthcomp\ and \borus\
normalisations are defined in the observed and cosmological frames, respectively, by
treating the cosmological redshift on of the primary continuum with a \zashift\
component; see \citealt{Walton22}). For the \relxilllpcp\ model, though, the situation is
more complicated as \relxilllpcp\ has a different definition for its normalisation than
\nthcomp. Here we therefore take the following approach: we perform an initial fit in
which both $C_{\rm{tor}}$ and the \borus\ normalisation are free to vary, then
determine the \nthcomp\ normalisation that would give the same primary continuum
as implied by the \relxilllpcp\ component (after again accounting for the cosmological
redshift with \zashift). We then calculate a range of allowed normalisations based on
this value and the fractional uncertainty on the \relxilllpcp\ normalisation (as seen by
an external observer) from the initial fit, and then repeat the fit with the \borus\
normalisation restricted to this range as our final version of the \relxilllpcp\ model.

Both models provide excellent fits to the hard X-ray data of \pg, with \chisq/DoF
= 1722/1645 and 1715/1645 for the \relxilllpcp\ and the \reflionx-based models,
respectively;
the best-fit parameters are given in Table \ref{tab_hard_param}. We note that if we set
the disc reflection fraction to zero the fit degrades by $\Delta\chi^{2} = 64$ and 71 for
4 fewer free parameters, further supporting the presence of relativistic reflection.
Both models show consistent solutions (although the \reflionx\ model allows for a
broader acceptable parameter space), and so we only show the relative contributions
of the different model components (and the quality of the fit via the data/model ratio)
for the \relxilllpcp\ model in Figure Fig. \ref{fig_hard_fit}). Even though the disc
reflection is well detected, only very loose constraints can be placed on the spin
parameter using only the hard X-ray data, with essentially the full range of prograde
values permitted at the 90\% level by both models. The inclination constraints
provided by both models are sensible for an unobscured AGN in the context of the
unified model (\eg\ \citealt{AGNunimod}). The best-fit reflection fractions also allow
reasonably standard values for a thin disc model, which should give \Rfrac\
$\sim$ 1.0--1.7 as long as there is no strong modification of the illumination of the
disc by gravitational lightbending, and also roughly solar values for the iron
abundance. Despite our efforts to determine $C_{\rm{tor}}$ for the distant
reflector, these values remain only poorly constrained.

%However, it is also
%important to
%note that the uncertainty range on this column does extend into the Compton-thick
%regime, so the uncertainty ranges on all of the other model parameters do allow for a
%contribution to the Compton hump from the distant reprocessor.

\subsection{Broadband Modelling}
\label{sec_broad}

Finally, given the bare nature of \pg, we also attempt to model the full broadband
\xmm+\nustar\ spectrum observed. Having established the presence of relativistic
reflection in the spectrum of \pg\ based on our analyses of the iron and hard X-ray
bandpasses, we focus on exploring whether this emission can also explain the soft
excess when the full \xmm+\nustar\ bandpass is considered. We therefore start by
re-fitting the reflection models constructed in Section \ref{sec_hard}, to the full
observed 0.3--78\,keV data, allowing for a few adjustments given the inclusion of
the soft X-ray data. First, we also allow the density of the disc to be a free
parameter in these fits, since the soft X-ray band is the portion of the X-ray
spectrum that is most sensitive to the density of the disc. As discussed in
\cite{Garcia16} the density parameter controls the level of free-free emission that
enters into the observed soft X-ray bandpass (see also \citealt{Jiang19agn,
Mallick22}). This free-free emission appears as a broad, thermal-like contribution
at low energies, and as the density of the disc increases more of this emission is
shifted up into the soft X-ray band. Second, with the soft X-ray data included we
also include the \ovii\ emission detected in the RGS data (see Section
\ref{sec_rgs}). We again use a single Gaussian to account for this feature, and
restrict the parameters of this component to line in the ranges found for the
equivalent model in our RGS analysis.

\begin{table*}
  \caption{Parameter constraints for the basic disc reflection models fit to
  the full broadband (\xmm+\nustar) X-ray data available for \pg. A * symbol
  indicates that the parameter was not free to vary in that particular model.}
\begin{center}
\vspace{-0.1cm}
\begin{tabular}{c c c c c c c}
\hline
\hline
\\[-0.2cm]
Model Component & \multicolumn{2}{c}{Parameter} & \relxilllpcp\ & \relxillcp\ & \relconvlp$\otimes$\reflionx & \relconv$\otimes$\reflionx \\
\\[-0.2cm]
\hline
\\[-0.2cm]
Primary Continuum & $\Gamma$ & & $2.29 \pm 0.01$ & $2.00 \pm 0.01$ & $1.82^{+0.01}_{-0.02}$ & $1.82^{+0.01}_{-0.04}$ \\ 
\\[-0.3cm]
(\nthcomp) & $kT_{\rm{e}}$\tmark[a] & [keV] & $>330$ & $>130$ & $69^{+25}_{-14}$ & $90^{+130}_{-30}$ \\
\\[-0.3cm]
& Norm\tmark[b] & [$10^{-3}$] & -- & -- & $1.28^{+0.08}_{-0.04}$ & $1.4 \pm 0.1$ \\
\\
\ovii\ & $E$ & [keV] & 0.569* & 0.569* & 0.569* & 0.569* \\
\\[-0.3cm]
emission\tmark[c] & $\sigma$ & [eV] & 5.4* & 5.4* & 5.4* & 5.4* \\
\\[-0.3cm]
& Norm & [$10^{-5}$] & 3.4* & 3.4* & 3.4* & 3.4* \\
\\
Disc Reflection & $a^*$ & & $0.95^{+0.01}_{-0.02}$ & $>0.99$ & $0.89 \pm 0.04$ & $0.77^{+0.06}_{-0.08}$ \\
\\[-0.3cm]
(various & $i$ & [\deg] & $40^{+1}_{-2}$ & $39^{+11}_{-3}$ & $<37$ & $<24$ \\
\\[-0.3cm]
combinations) & $h$ & [\rg] & $2.2^{+0.3}_{-0.1}$ & -- & $<2.5$ & -- \\
\\[-0.3cm]
& $q_{\rm{in}}$ & & -- & $6.4 \pm 0.5$ & -- & $4.9^{+1.1}_{-1.0}$ \\
\\[-0.3cm]
& $R_{\rm{br}}$ & \rg\ & -- & $7.0^{+2.7}_{-1.2}$ & -- & $10.0^{+3.2}_{-1.9}$ \\
\\[-0.3cm]
& $q_{\rm{out}}$ & & -- & 3.0* & -- & 3.0* \\
\\[-0.3cm]
& $A_{\rm{Fe}}$ & [solar] & $3.8^{+0.3}_{-0.5}$ & $1.7^{+0.2}_{-0.3}$ & $0.76 \pm 0.04$ & $0.79^{+0.05}_{-0.03}$ \\
\\[-0.3cm]
& $\log\xi$ & $\log$[\ergcmps] & $<0.1$& $2.71^{+0.04}_{-0.10}$ & $3.01^{+0.05}_{-0.03}$ & $2.97^{+0.06}_{-0.03}$ \\
\\[-0.3cm]
& $\log n$ & $\log$[\pcmcub] & $17.0^{+0.1}_{-0.2}$ & $17.2 \pm 0.1$ & $17.98 \pm 0.05$ & $18.0^{+0.1}_{-0.2}$ \\
\\[-0.3cm]
& \Rfrac\tmark[d] & & $7.5 \pm 0.3$ & $3.5^{+0.5}_{-0.4}$ & $1.0^{+0.5}_{-0.1}$ & $1.1^{+0.1}_{-0.5}$ \\
\\[-0.3cm]
& Norm & & $6.1^{+3.3}_{-0.9} \times 10^{-4}$ & $1.9^{+0.3}_{-0.1} \times 10^{-5}$ & $0.115^{+0.010}_{-0.005}$ & $0.11 \pm 0.01$ \\
\\
Distant Reflection & $\log$\,\nh\ & log[\pcmsq] & $24.7 \pm 0.2$ & $>24.7$ & $23.8^{+0.2}_{-0.1}$ & $23.8^{+0.2}_{-0.1}$ \\
\\[-0.3cm]
(\borus) & $C_{\rm{tor}}$ & & $0.76 \pm 0.02$ & $0.76^{+0.04}_{-0.13}$ & $0.69^{+0.20}_{-0.11}$ & $0.80^{+0.08}_{-0.10}$ \\
\\[-0.3cm]
& Norm & [$10^{-3}$] & $4.0^{+0.4}_{-0.5}$ & $3.4^{+0.5}_{-0.6}$ & -- & -- \\
\\[-0.2cm]
\hline
\\[-0.25cm]
\chisq/DoF & & & 2366/2046 & 2269/2045 & 2275/2047 & 2244/2046 \\
\\[-0.25cm]
\hline
\hline
\end{tabular}
\label{tab_broad1}
\end{center}
\flushleft
$^a$ For the lamppost variants $kT_{\rm{e}}$ is quoted in the rest-frame of the X-ray
source (\ie prior to any gravitational redshift), based on the best-fit lamppost geometry. \\
$^b$ The primary \nthcomp\ continuum is incorporated within the \relxilllpcp\ model,
and so does not have a separate normalisation. \\
$^c$ These parameters are fixed, based on the analysis of the RGS data (see Section
\ref{sec_rgs}). The line energy is given in the
rest-frame of \pg. \\
$^d$ \Rfrac\ is not formally a free parameter for the \reflionx-based models, but is
again calculated following the method outlined in \cite{Walton13spin}, i.e. via
a comparison with the \pexrav\ model (\citealt{pexrav}). \\
\vspace{0.4cm}
\end{table*}

The \relxilllpcp\ model does not provide a particularly good fit to the broadband
data, even if we relax any attempt to set the \borus\ normalisation based on the
\relxilllpcp\ parameters (\chisq/DoF = 2366/2046). While this model can fit the
data below $\sim$30\,keV well, at higher energies the model does not reproduce
the overall shape of the observed spectrum, leaving a clear excess of emission
at the highest energies probed by \nustar\ (see Figure \ref{fig_broad_ratio}, left
panels). It is possible that part of the issue here is that the lamppost geometry
assumed for the corona is overly simplistic for these sources. We therefore also
fit a version of this model that instead adopts a phenomenological broken
powerlaw for the emissivity profile -- and thus is agnostic about the exact
geometry of the corona -- using the \relxillcp\ model instead. Here, instead of
being set solely by the height of the corona, the emissivity is described by two
indices ($q_{\rm{in}}$ and $q_{\rm{out}}$) separated by a break radius
($R_{\rm{br}}$), where over a given radial regime the emissivity has the form
$\epsilon(r) \propto r^{-q}$. The inner emissivity index and the break radius are
free to vary in this model, while the outer index is fixed to $q_{\rm{out}} = 3$ (as
expected for large radii; \citealt{Reynolds97feK}). This model improves both the
overall fit (\chisq/DoF = 2269/2045) and reduces the strength of the hard
excess, but it is not able to remove it completely. Similar results have been
reported for Ark\,120 -- another well-known bare AGN -- for this model by
\cite{Porquet18}, who also attempted to apply similar \xillver-based reflection
models to broadband data (also \xmm+\nustar\ in coordination) and found they
did not provide a satisfactory fit across the full bandpass (see also
\citealt{Matt14}).

\begin{figure}
\begin{center}
\hspace*{-0.35cm}
\rotatebox{0}{
{\includegraphics[width=235pt]{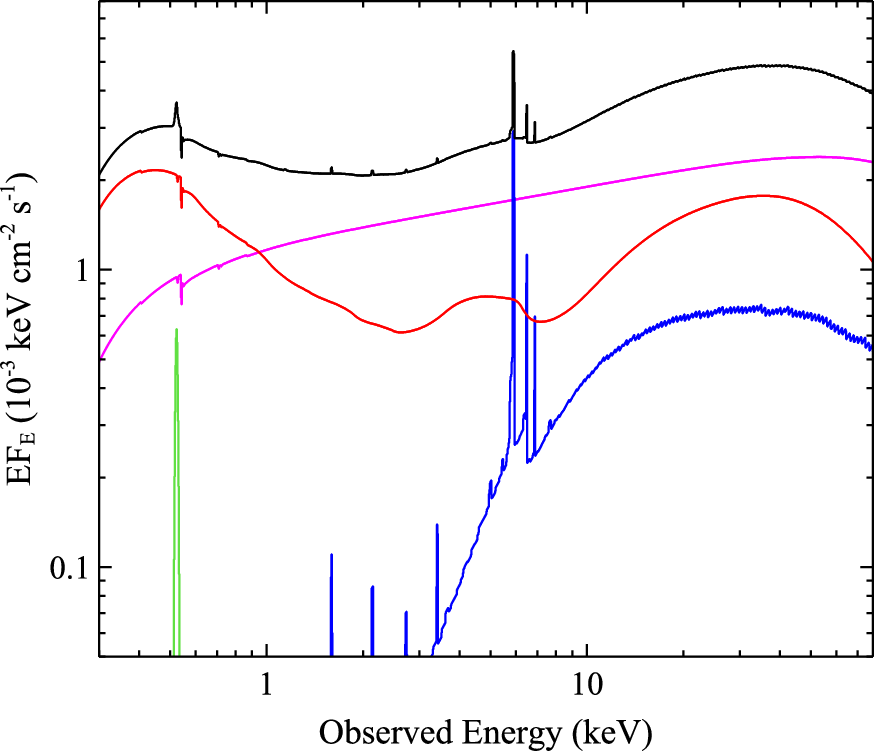}}
}
\end{center}
\vspace*{-0.3cm}
\caption{The best-fit \reflionx-based disc reflection model with a broken
powerlaw emissivity profile, applied to the full \xmm+\nustar\ broadband dataset.
The colours are the same as Figure \ref{fig_hard_fit} (top panel), with the
narrow \ovii\ emission additionally shown in green.
}
\label{fig_broad_mod}
\end{figure}

In contrast, \reflionx-based disc reflection models have no problem reproducing
the broadband data. We fit both of the variants considered for the \xillver-based
models (lamppost geometry, broken powerlaw emissivity), and both can
simultaneously fit the soft excess, the broad iron line and the Compton hump
(see Figure \ref{fig_broad_ratio}, right panels), although the broken powerlaw
emissivity does again provide the better statistical fit (the lamppost and broken
powerlaw emissivities gives fits with \chisq/DoF = 2311/2047 and 2244/2046,
respectively). The parameter constraints for all four of these `basic' reflection
models when applied to the broadband data are given in Table
\ref{tab_broad1}.\footnote{Note that since neither of the \relxill-based models
reported here provide satisfactory fits to the broadband data with the initial fits
in which both the \borus\ normalisation and $C_{\rm{tor}}$ are free parameters,
we did not undertake the subsequent step discussed for the \relxilllpcp\ fit to
the hard X-ray data (Section \ref{sec_hard}) in which the \borus\ normalisation
is further limited to the range implied by the \relxill\ primary continuum. Instead,
we simply report the parameter constraints for these initial fits for the
\relxill-based models. In contrast, since the primary continuum and the disc
reflection are separate model components in the \reflionx-based models, the
\borus\ normalisation remains linked to that of the \nthcomp\ components in
these fits, and so is once again not a separate parameter to report.} We show
the best-fit \reflionx-based model with the broken powerlaw emissivity profile
in Figure \ref{fig_broad_mod}.

A key difference between these \xillver-based and \reflionx-based models lies in the
densities they return for the accretion disc, with the \xillver-based models implying
\logne\ $\sim 17$ while the \reflionx-based models implying \logne\ $\sim 18$. As
such, within the observed bandpass these \xillver-based models include a smaller
contribution from the free-free emission that increasingly dominates the soft X-ray
reflection continuum at higher densities (\citealt{Garcia16}). In turn, the relative
contribution from reflection in the soft band is therefore much higher for these
\reflionx-based models. Because of the weaker soft excesses in these \xillver-based
models they pivot the slope of the primary continuum to steeper values in order to
successfully reproduce the soft X-ray data, as this part of the spectrum dominates
the overall photon statistics of the broadband data. This pivoting happens most
severely for the lamppost variant. In turn, this drives both the reflection
fraction and the iron abundance up, as the model tries as best it can to use this
emission to reproduce the highest energy data seen by \nustar\ and softer ionising
spectra tend to produce reflection spectra with weaker iron lines. In the case of
the lamppost variant this also drives the ionisation parameter down (as doing so
results in a harder reflection spectrum). Ultimately, though, the steeper photon indices
mean that the reflected emission is still too steep to match the high energy data,
resulting in the hard excesses seen in Figure \ref{fig_broad_ratio}. Because the
\reflionx-based models are able to provide stronger soft excesses, no such pivoting
of the continuum is required, and so they are able to better reproduce both the soft
excess and the hard X-ray data.

The two \reflionx-based models that do reproduce the full bandpass well occupy
broadly similar areas of parameter space, and in the case of the lamppost model
(where the most direct comparison can be made) the results are generally
consistent with those found for the equivalent fits to the hard X-ray data (though
with tighter parameter constraints now, related to the higher total S/N in the full,
broadband dataset that comes from the inclusion of the \xmm\ data below 2\,keV).
With regards to the disc reflection specifically, both imply relatively high but not
extreme spins, and relatively normal reflection fractions (\Rfrac\ $\sim $ 1;
\citealt{pexrav, Walton13spin}). The best-fit inclinations ($i \lesssim 25$\deg)
suggest that we are viewing the disc almost face-on; again, this is consistent with
the general expectation for an AGN with such low obscuration based on the
unified model (\citealt{AGNunimod}). Finally, both models also imply very similar
densities for the accretion disc of $n \sim 10^{18}$\,cm$^{-3}$, in excess of the
value traditionally adopted in most older generations of reflection model (which
generally assumed $10^{15}$\,cm$^{-3}$; \eg\ \citealt{reflion, xillver}).

\begin{figure}
\begin{center}
\hspace*{-0.35cm}
\rotatebox{0}{
{\includegraphics[width=235pt]{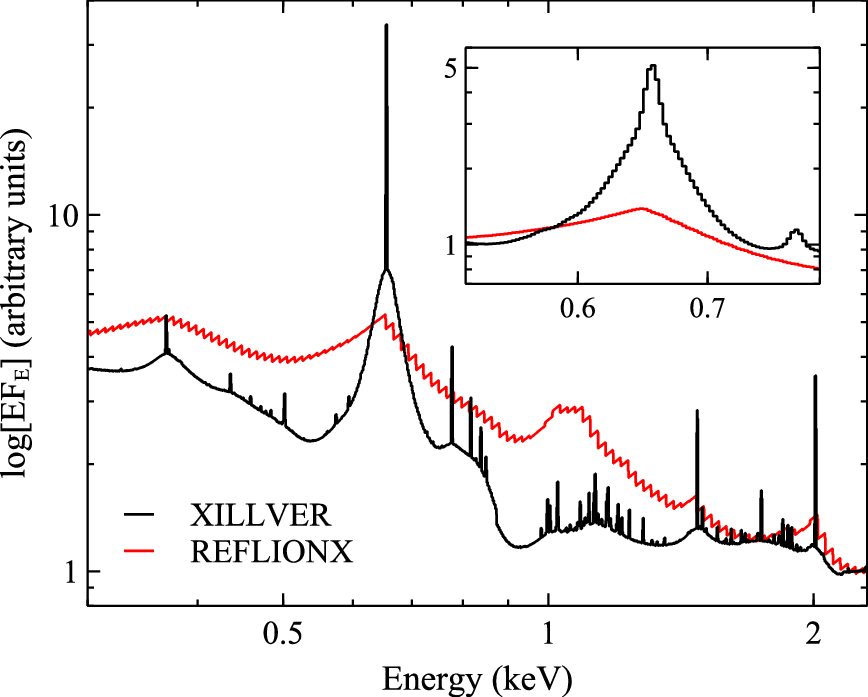}}
}
\end{center}
\vspace*{-0.3cm}
\caption{\textit{Main panel:} A comparison of the rest-frame \xillver\ (black) and
\reflionx\ (red) reflection models in the soft band for a density of $n = 10^{18}$
cm$^{-3}$ and an ionisation parameter of $\xi = 10^{3}$\,\ergcmps\ (\ie\ no
relativistic blurring has been applied). We also assumed an \nthcomp\ ionising
continuum with $\Gamma = 1.8$ and $kT_{\rm{e}} = 60$\,keV, as well as a solar
iron abundance for both models, and the models have been normalised to give
the same flux at 2.25\,keV. The \xillver\ model appears to have stronger oxygen
emission at $\sim$0.65\,keV for this parameter combination than the \reflionx\
model. Note also that the \reflionx\ model has a coarser internal spectral
resolution, resulting in the visible saw-toothing in the main panel. \textit{Inset:}
The same models smoothed with a Gaussian of width 25\,eV (so that they are
now plotted with broadly the same resolution) and zoomed-in on the oxygen
emission in question. The models are also now normalised so that the averages
of the fluxes at 0.55 and 0.75\,keV are the same in both cases. The quantitative
comparisons discussed in the main text confirm the stronger \oviii\ emission in
\xillver.
}
\label{fig_refcomp}
\end{figure}

\subsubsection{Xillver: The Importance of Oxygen}
\label{sec_ox}

In principle it should still be possible for \xillver\ to provide a low-energy reflection
continuum with a similar strength as \reflionx\ by moving to higher densities than the
fits in Table \ref{tab_broad1} imply. However, a key reason that the \relxillcp\ model
(which comes closest of the \xillver-based models considered here to fitting the
broadband data) does not do this is that at these higher densities than reported in
Table \ref{tab_broad1} \xillver\ predicts the presence of an extremely strong \oviii\
emission line in the rest-frame reflection spectrum at the ionisation parameters
preferred by this model (and also the \reflionx-based fits; $\xi \sim 1000$~\ergcmps).
This results in a structure to the soft excess (even after the application of the
relativistic blurring) that the data do not want. The corresponding \oviii\ emission in
the \reflionx\ model is much weaker (see Figure \ref{fig_refcomp}), which also helps
this model fit the broadband data successfully. When compared against a local
powerlaw continuum defined between the energies 0.55 and 0.75 keV, the equivalent
width of the oxygen emission in the \xillver\ model is a factor of $\sim$3 larger than
in \reflionx. This difference in EW is in part because of the stronger soft X-ray
continuum seen in \reflionx\ for a given density, but also because the absolute \oviii\
line flux (evaluated as the flux difference between the reflection models and the
local powerlaw continua described earlier in the 0.55-0.75\,keV band) is a factor of
$\sim$2 larger in the \xillvercp\ model for these parameters when the two models
are normalised to have equal fluxes at 2.25\,keV (away from the main impact of the
different low-energy reflection continua in the two models and avoiding any obvious
line emission)\footnote{We also find a similar difference if we normalise the \reflionx\
and \xillver\ models to have the same flux over broader bandpasses of 0.1--10\,keV
and 0.1--1000\,keV, instead of at 2.25\,keV specifically.}. The difference in these
\oviii\ line fluxes cannot obviously be related to the different elemental abundances
the two models adopt, as their oxygen abundances (relative to Hydrogen) only differ
by 10\%.

% Grevesse & Sauval: O/Fe = 21.38
% Morrison & McCammon: O/Fe = 22.39

\begin{table}
  \caption{Parameter constraints for the \xillver-based disc reflection model where
  the flux of the \oviii\ emission in the \xillver\ component is allowed to be reduced
  by an additional \gabs\ component, fit to the full broadband (\xmm+\nustar) X-ray
  data. A * symbol indicates that the parameter was not free to vary.}
\begin{center}
\vspace{-0.1cm}
\begin{tabular}{c c c c c c c}
\hline
\hline
\\[-0.2cm]
Model Component & \multicolumn{3}{c}{Parameter} & \\
\\[-0.2cm]
\hline
\\[-0.2cm]
Primary Cont. & $\Gamma$ & & $1.92 \pm 0.01$ \\ 
\\[-0.3cm]
(\nthcomp) & $kT_{\rm{e}}$ & [keV] & $>130$ \\
\\[-0.3cm]
& Norm & [$10^{-3}$] & $1.49^{+0.10}_{-0.03}$ \\
\\
\ovii\ & $E$ & [keV] & 0.569* \\
\\[-0.3cm]
emission\tmark[a] & $\sigma$ & [eV] & 5.4* \\
\\[-0.3cm]
& Norm & [$10^{-5}$] & 3.4* \\
\\
\relconv\ & $a^*$ & & $0.95 \pm 0.01$ \\
\\[-0.3cm]
& $i$ & [\deg] & $42 \pm 2$ \\
\\[-0.3cm]
& $q_{\rm{in}}$ & & $7.0^{+0.4}_{-0.5}$ \\
\\[-0.3cm]
& $R_{\rm{br}}$ & \rg\ & $6.0^{+1.0}_{-0.5}$ \\
\\[-0.3cm]
& $q_{\rm{out}}$ & & 3.0* \\
\\
\gabs\tmark[b] & $E$ & [keV] & 0.654* \\
\\[-0.3cm]
& $\sigma$ & [eV] & 1* \\
\\[-0.3cm]
& Strength & [$10^{4}$] & $>1.5$ \\
\\
\xillvercp\ & $A_{\rm{Fe}}$ & [solar] & $1.8^{+0.2}_{-0.1}$ \\
\\[-0.3cm]
& $\log\xi$ & $\log$[\ergcmps] & $2.70^{+0.01}_{-0.05}$ \\
\\[-0.3cm]
& $\log n$ & $\log$[\pcmcub] & $18.00^{+0.02}_{-0.08}$ \\
\\[-0.3cm]
& \Rfrac\tmark[c] & & $1.8^{+0.3}_{-0.1}$ \\
\\[-0.3cm]
& Norm & [$10^{-5}$] & $4.5^{+0.6}_{-0.3}$ \\
\\
Distant Refl. & $\log$\,\nh\ & log[\pcmsq] & $>24.7$ \\
\\[-0.3cm]
(\borus) & $C_{\rm{tor}}$ & & $0.6^{+0.05}_{-0.10}$ \\
\\[-0.2cm]
\hline
\\[-0.25cm]
\chisq/DoF & & & 2247/2045  \\
\\[-0.25cm]
\hline
\hline
\end{tabular}
\label{tab_broad2}
\end{center}
\flushleft
$^a$ As before, these parameters are fixed based on the analysis of the RGS data (see
Section \ref{sec_rgs}), and the line energy is given in the rest-frame of \pg. \\
$^b$ The \gabs\ component is applied to the \xillvercp\ component prior to the further
application of \relconv, so that the overall disc reflection component here has the form
\relconv $\otimes ($\gabs$\times$\xillvercp$)$. Again the line energy is given in the
rest frame of \pg. \\
$^c$ \Rfrac\ is not formally a free parameter here, now that the \relconv\ and \xillver\
components are separated, and but is again calculated following the method outlined
in \cite{Walton13spin}, i.e. via a comparison with the \pexrav\ model (\citealt{pexrav}). \\
\vspace{0.4cm}
\end{table}

\begin{figure}
\begin{center}
\hspace*{-0.35cm}
\rotatebox{0}{
{\includegraphics[width=235pt]{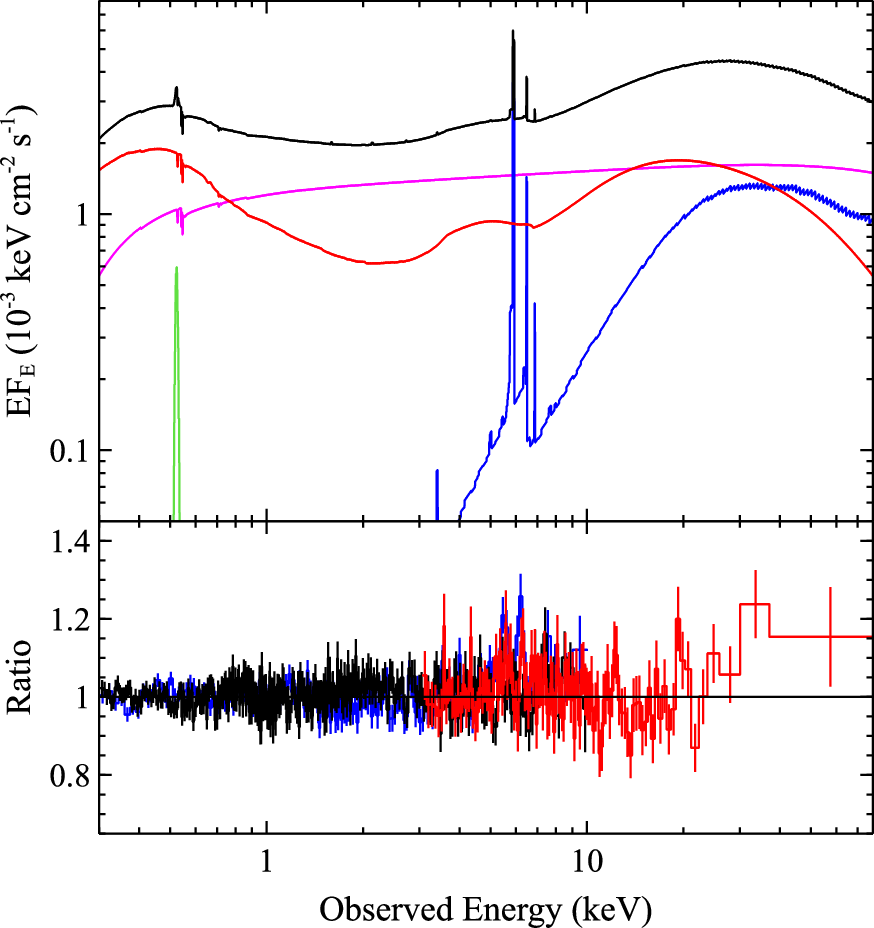}}
}
\end{center}
\vspace*{-0.3cm}
\caption{\textit{Top panel:} the best-fit for the \xillver-based disc reflection model
where the \oviii\ emission in the \xillver\ component is allowed to be modified by an
additional \gabs\ component to mimic a sub-solar oxygen abundance, applied to the
full \xmm+\nustar\ broadband dataset. The colours in both panels have the same
meaning as Figures \ref{fig_hard_fit} and \ref{fig_broad_mod}, and the data have
again been rebinned for visual clarity.
}
\label{fig_broad_fit2}
\end{figure}

The above comparison between \reflionx\ and \xillver\ suggests that being able to
adjust the strength of the oxygen emission may be an important factor in
determining whether \xillver-based models can fit broadband X-ray data from AGN
successfully (and particularly the soft excess, when there is a clear view of this
feature). Most reflection models allow for the iron abundance to be a free parameter,
given the importance of the relativistically broadened iron line, but it is rare for
reflection models to allow for non-solar oxygen abundances as well. Indeed, there is
currently no version of the \xillver\ model that is both suitable for use with AGN and
has the oxygen abundance as a free parameter which would allow us to test this
possibility directly. We therefore attempt to mimic this possibility by applying a
Gaussian absorption line to the intrinsic reflection spectrum (using a \gabs\
component) in order to remove some of the flux from the oxygen emission prior to
the application of the relativistic blurring (since in this case it appears that the
\xillver\ model has more oxygen emission than the data wants). Doing so requires
separate treatment of the reflection and the relativistic blurring, and so here we
replace the \relxillcp\ component with the combination \nthcomp\ $+$ \relconv
$\otimes [$\gabs$\times$\xillvercp$]$, where the \xillvercp\ component is configured
to only provide reflected emission. The energy of the \gabs\ component is fixed at
0.654\,keV in the rest frame of \pg, and the width of the line is fixed at 1\,eV, so the
only extra free parameter is the strength of the absorption. With this model setup, we
can once again directly link the \borus\ normalisation to that of the primary \nthcomp\ 
continuum. Other aspects of the model setup remain the same as the equivalent
model discussed in the previous section.

This model combination provides a better fit to the broadband data than the basic
implementation discussed previously (\chisq/DoF = 2247/2045), and critically
there is no longer a significant excess of emission at high energies. This is because
the reduction in the \oviii\ flux means the reflection component is now more easily
able to account for the soft excess by increasing the density of the disc (which is
now also $n \sim 10^{18}$\,\pcmcub, similar to the \reflionx-based fits), and
subsequently the powerlaw continuum no longer needs to pivot to softer values to
the same degree in order to fit this part of the spectrum. The best-fit model is
shown in Figure \ref{fig_broad_fit2}, along with the data/model ratio, and the full
set of parameter constraints for this model are given in Table \ref{tab_broad2}. With
the best-fit parameter combination, the \gabs\ component removes $\sim$half of
the total \oviii\ flux in the \xillvercp\ component, again assessed by comparing the
\xillvercp\ model against a local powerlaw continuum defined between 0.55 and
0.75\,keV both with and without the \gabs\ component included. This may imply
that an oxygen abundance of $A_{\rm{O}}$/solar $\sim$ 0.5 would be preferred by
this model, although it is also worth noting that the remaining \oviii\ flux would still
imply that within the reflection model itself this feature has an equivalent width
comparable to that of the \reflionx\ model with a solar oxygen abundance.
The mild remaining high-energy residuals will likely be accounted for with future
variants of \relxill/\xillver\ that will incorporate the improved treatment of Compton
scattering discussed in \cite{Garcia20}, as this will result in reflection spectra with
slightly higher high-energy fluxes (though we stress that this by itself would not be
sufficient to account for the excesses seen in Figure \ref{fig_broad_ratio} without
still needing to modify the strength of the oxygen emission).

We also note that a similar improvement is seen when allowing the \oviii\ line
strength to vary for the lamppost model as well (i.e. treating the relativistic
reflection as \relconvlp $\otimes [$\gabs$\times$\xillvercp$]$), though as with the
\reflionx-based models the broken-powerlaw emissivity once again provides the
superior fit (the fit for the lamppost model improves to \chisq/DoF = 2287/2046).
We do not present this model in full, as the best-fit solution is broadly similar to
that already presented in Table \ref{tab_broad2}, but note that the lamppost
variant once again prefers a slightly higher spin parameter ($a^* > 0.99$), similar
to the results seen from the \reflionx-based\ models (Table \ref{tab_broad1}), and
the constraint on the height of the corona is broadly similar to that found for the
\reflionx-based lamppost variant ($h \lesssim 2$\,\rg).

\section{Discussion}
\label{sec_dis}

\pg\ is a source of notable interest, as it has the largest reverberation-mapped mass
reported for any AGN to date: $\log[m_{\rm{BH}}] = 9.01^{+0.11}_{-0.16}$
(\citealt{Kaspi00, Peterson04}). Furthermore, the high-resolution RGS data reported
here confirm its nature as one of the relatively rare examples of a genuinely bare AGN
(Section \ref{sec_rgs}).

\subsection{The Soft Excess}

The bare nature of \pg\ offers an opportunity to explore the nature of the soft excess
-- which is very clearly present in the spectrum of \pg\ (Figure \ref{fig_spec}) -- via
broadband X-ray observations taken with \xmm\ and \nustar. Together, the
observations presented here provide simultaneous coverage over the 0.3--78\,keV
bandpass. Currently, two leading models are typically considered as the potential
origins of this emission in the literature: relativistic reflection from the accretion disc,
and the presence of a second, `warm' corona (in addition to the `hot' corona that
gives rise to the primary X-ray continuum seen above 2\,keV).

We have focused our work on the question of whether the relativistic reflection model
can successfully fit the broadband dataset. In this model, the soft excess comes about
through the combination of the low-energy emission lines predicted by ionised
reflection and the appreciable free-free emission expected from dense, ionised plasma;
the strong relativistic effects close to the black hole broaden and blend these features
into a relatively smooth emission profile (\eg\ \citealt{Crummy06, Walton13spin,
Jiang19agn, Mallick22}). We focus on this issue because, as a potential part of the
reflection component from the inner accretion disc, simultaneous broadband
observations such as those presented here represent the key stress test of whether
this model remains plausible, and there are contrasting claims in the literature with
regards to whether it can successfully fit the broadband data obtained for other bare
AGN (\eg\ \citealt{Porquet18, Garcia19, Ursini20wc, Xu21, Pottayil24}). It is worth
stressing that, regardless of the nature of the soft excess, the combined \xmm+\nustar\
data considered here independently confirm the presence of relativistic reflection in
the spectrum of \pg: a relativistically broadened iron emission line is clearly detected
(Section \ref{sec_fek}), and the corresponding high-energy reflected continuum is also
observed (Figure \ref{fig_spec}).

Although we have not explicitly considered the warm corona model in this work,
this is simply because such fits are already presented in a separate analysis of these
data, included as part of an ensemble analysis of a sample of AGN with broadband
spectra (\citealt{Mallick25}). That work initially attempts to model the soft excess with
relativistic reflection using \xillver-based reflection models, finding the same issues
with these fits for \pg\ as discussed in Section \ref{sec_broad} (and similar to prior
studies of a few other sources; \citealt{Porquet18, Porquet24, Ursini20wc}).
\cite{Mallick25} subsequently include a warm corona to account for the soft excess in
\pg, finding that this interpretation is able to successfully fit the broadband data.
We therefore stress that we are not attempting to argue that this is not a viable model
for the soft excess by focusing only on the reflection case.

As part of our focus on the reflection interpretation, we have chosen to test both of the
leading reflection models commonly used in the recent literature: \xillver\ and \reflionx.
Both of these models allow for ionised reflection, and their latest versions both also
include the density of the disc as a free parameter. This is a key parameter that controls
the free-free emission seen in these models at lower energies (\citealt{Garcia16}), and
is therefore an important factor to consider when attempting to model the soft excess
with relativistic reflection (as already highlighted by several previous authors; \eg\
\citealt{Jiang19agn, Mallick22}). 

Ultimately we find that both reflection models are able to reproduce the broadband
\xmm+\nustar\ data observed, including the soft excess. However, we do find an issue
when initially attempting to fit the data with the \xillver-based model, which relates to
the very strong \oviii\ emission line predicted by this model at higher densities (much
stronger than the equivalent \reflionx\ model, see Section \ref{sec_ox}). The
strength of this line is sufficient that it introduces structure into the model that the
data do not want.  Allowing a weaker \oviii\ emission line is required to enable the
\xillver-based model to successfully reproduce the broadband data.  This same issue
is likely what resulted in \cite{Porquet18, Porquet24} and \cite{Ursini20} concluding
the relativistic reflection cannot reproduce both the soft excess and the hard X-ray
data in other bare AGN with coordinated \xmm\ + \nustar\ observations even when
treating the disc density as a free parameter. Those analyses all used \xillver-based
reflection models, and the resulting fits leave residuals that are all qualitatively similar
to those we find for the \xillver-based models for \pg\ without any modification to the
oxygen emission included (Figure \ref{fig_broad_ratio}). No such correction is required
for the \reflionx-based models, since \reflionx\ typically predicts both a stronger
low-energy reflection continuum and weaker \oviii\ emission.

In principle, there are several potential reasons that the \oviii\ line from the disc in
\pg\ could be weaker than the \oviii\ line predicted in the \xillver\ model. An obvious
astrophysical possibility is that the oxygen abundance in \pg\ is below the solar
value.  Non-solar iron abundances have been reported for several AGN even when
density of the disc is included as a free parameter (\eg, \citealt{Mallick18, Jiang19agn,
Jiang22iras,Walton21eso}), so non-solar oxygen abundances may not be
unreasonable. However, while this may be plausible if the \xillver\ model is considered
in isolation, this cannot obviously explain the differences between the \xillver\ and
\reflionx\ models in terms of their \oviii\ emission, as the two models adopt similar
values for the solar abundance of oxygen.  The detection of the narrow \ovii\ emission
in the RGS spectrum (Section \ref{sec_rgs}) may also argue against this possibility. A
more critical issue may be that the \oviii\ line is a resonant line, and the radiative
transfer of such emission lines is highly non-trivial, relating in particular to the
re-absorption and scattering of the line photons as they emerge from the disc. The
two models take different approaches to dealing with these lines, which likely explains
why the two models predict different line strengths; \reflionx\ treats these lines using
an escape fraction approach (\citealt{reflion}), while \xillver\ assumes these lines
follow the same radiative transfer as the continuum emission, which is solved via the
Feautrier method (\citealt{xillver}).  As such, \xillver\ does not currently account for
any additional line opacity when the lines themselves have appreciable optical depths,
a potential issue for the strength of the \oviii\ emission (though not necessarily for
other key lines such as Fe K).\footnote{In more detail: \xillver\ uses the \xstar\ routines
to compute continuum and line emissivities (and opacities). By default, line emissivities
are calculated assuming an optically-thin medium, in which case the line emissivity
scales linearly with the ion fraction and excitation rate. However, in optically-thick
conditions, photons are resonantly trapped and scattered, in which case the escape
probability of a line scales as the inverse of the optical depth. This dependency is not
currently taking into account in the \xillver\ models, which can lead to an
overestimation of the \oviii\ emission. Note, however, this effect does not affect the
Fe K emission, as those are fluorescent lines involving photoionization and
recombination rather bound-bound transitions.} Should \xillver\ end up
over-predicting the \oviii\ emission, as is likely at least to some extent, one would
need to remove some of this flux in order to produce realistic reflection spectra.
Further investigation of this issue will be important to pursue in the future, but is
beyond the scope of this paper.

\subsection{Density of the Accretion Disc}

Considering the different models that successfully fit the data, we find an overall
constraint on the density of the disc of \logne\ = \neresult, significantly in excess of
the typical density of \logne\ = 15 assumed in most prior reflection models. The only
prior constraint available in the literature for \pg\ comes from \cite{Jiang19agn}, who
find \logne\ $< 15.9$,  which is rather different to the result found here. It is worth
noting, though, that this prior constraint is derived from a short snapshot observation
taken with \xmm, only providing spectral coverage up to 10\,keV. These limitations
meant \cite{Jiang19agn} ended up fixing several of the reflection parameters to
assumed values (\eg\ they assumed a powerlaw emissivity with $q=3$), since they
were otherwise unconstrained. As such the deeper, broadband data considered here
likely provide more reliable constraints on the density of the disc.

\begin{figure}
\begin{center}
\hspace*{-0.35cm}
\rotatebox{0}{
{\includegraphics[width=235pt]{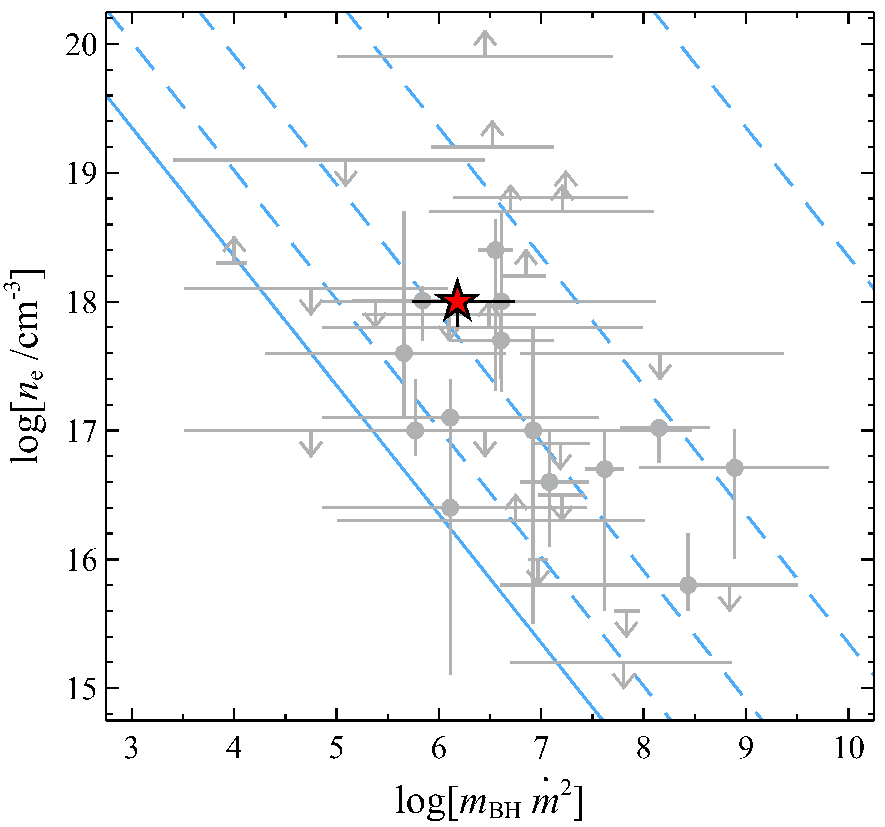}}
}
\end{center}
\vspace*{-0.3cm}
\caption{The position of \pg\ in the \logne\ vs $\log [m_{\rm{BH}} \dot{m}^2]$
plane, compared to other AGN for which reflection-based density constraints for
the inner disc are available.  Our results for \pg\ are indicated by the red star, while
the broader AGN population are shown in the background as grey points (compiled
from: \citealt{Jiang19agn, Jiang19_1h0419, Jiang22iras, Mallick18,Mallick22,
Garcia19,Walton21eso, Xu21,Wilkins22,Pottayil24}, 2025, \textit{in preparation}).  Note
that the prior constraint for \pg\ presented in \citet{Jiang19agn} is not included as
part of this comparison sample., nor are the constraints for Ark\,564 or PG\,1229+204
from that work (the other two sources for which UV data were used when estimating
$\log [m_{\rm{BH}} \dot{m}^2]$; see footnote \ref{ft_UV}). The results for \pg\ sit
within the scatter of the loose anti-correlation between \logne\ and $\log
[m_{\rm{BH}} \dot{m}^2]$ exhibited by these other sources.  The solid blue line
shows the predicted \logne\ $\propto -\log [m_{\rm{BH}} \dot{m}^2]$ trend
predicted by \citet{Svensson94} for a radius just outside the inner radius of the disc
(formally we assume $R = 4 R_{\rm{G}}$ and $R_{\rm{in}} = 2 R_{\rm{G}}$, following
\citealt{Jiang19agn}) and the limiting case where no accretion power is dissipated in
the corona (\ie\ $f = 0$). The dashed blue lines show how this anti-correlation is
expected to vary as $f$ increases (from left to right: $f = 0.4, 0.7, 0.9, 0.99$).
Allowing for a reasonable range in $f$ across different systems, the data show a
good overall agreement with the \citet{Svensson94} model.
}
\label{fig_density}
\end{figure}

In the disc--corona model of \cite{Svensson94} the density of the inner disc is
expected to be inversely proportional to the product $m_{\rm{BH}} \dot{m}^2
(1-f)^3$, where $\dot{m}$ is the Eddington-scaled accretion rate (\ie\ $\dot{m} = 
\dot{M}c^2/L_{\rm{E}}$) and $f$ is the fraction of the accretion power dissipated in
the corona, assuming the disc is dominated by radiation pressure. In order to place
the results for \pg\ in the context of this model, and the growing ensemble of
reflection-based density constraints for AGN accretion discs (compiled from
\citealt{Jiang19agn, Jiang19_1h0419, Jiang22iras, Mallick18,Mallick22,
Garcia19,Walton21eso,Xu21,Wilkins22,Pottayil24}),  we show where \pg\ lies in the
\logne\ vs $\log [m_{\rm{BH}} \dot{m}^2]$ plane in Figure \ref{fig_density} (omitting
the prior density constraint presented in \citealt{Jiang19agn} for \pg\ from the
contextual data plotted to ensure this source does not appear twice). To facilitate a
self-consistent comparison between our updated result for \pg\ and this broader
AGN sample, following \cite{Jiang19agn} we determine the value of $\log
[m_{\rm{BH}} \dot{m}^2]$ by combining the reverberation-mapped mass
constraint with the accretion rate estimated from the optical luminosity using the
method outlined in \cite{Raimundo12}:

\vspace{-0.2cm}
\begin{equation}
\frac{\dot{M}}{M_{\odot} ~ \rm{yr}^{-1}} = 1.53 \bigg( \frac{\nu L_{\nu}}{10^{45} ~ \cos i} \bigg)^{3/2} \bigg( \frac{m_{\rm{BH}}}{10^{8}} \bigg)^{-1} \bigg( \frac{\nu}{\nu_{4392}} \bigg)^{-2}
\label{eqn_mdot}
\end{equation}
\vspace{0.05cm}

\noindent{Here},  the monochromatic luminosity $\nu L_{\nu}$ is measured in units
of \ergps, $\nu$ is the photon frequency at which this luminosity is measured and
$\nu_{4372}$ is the photon frequency that corresponds to the wavelength of
4392\AA\ for which the above relation is calibrated. We calculate $\dot{M}$ using
the 5100\AA\ luminosity reported by \cite{Peterson04} here, as this is closer in
wavelength to 4392\AA\ than the UVM2 band used for \pg\ in \cite{Jiang19agn}.
\footnote{\label{ft_UV} We also note that \cite{Jiang19agn} did not properly
account for the frequency term included in equation \ref{eqn_mdot}, the impact of
which is most significant for the small number of sources where UV bands were
used to estimate $\dot{M}$ in that work. Use of the UV bands without making this
correction, as was the case for \pg, significantly overpredicts $\dot{m}$, and thus
the product $m_{\rm{BH}} \dot{m}^2$.} This combination gives $\log [m_{\rm{BH}}
\dot{m}^2] = 6.18^{+0.56}_{-0.44}$.

The ensemble of AGN constraints shown in Figure \ref{fig_density} are broadly
consistent with the anti-correlation expected in the \logne\ vs $\log [m_{\rm{BH}}
\dot{m}^2]$ plane based on the \cite{Svensson94} model, though there is
significant scatter (potentially related to different AGN exhibiting different values
of $f$; \citealt{Mallick25}).  Although the relatively large density inferred for \pg\ may
initially seem surprising given its large mass, it actually sits comfortably within the
scatter seen from other AGN in this plane. \pg\ is therefore also broadly consistent
with the same overall anti-correlation between \logne\ and $\log [m_{\rm{BH}}
\dot{m}^2]$, and thus the general prediction of \cite{Svensson94}.

\subsection{Black Hole Spin}

The large mass of the black hole in \pg\ makes it a key target for expanding the
sample of AGN spin measurements. The mass--spin plane is a key diagnostic for
SMBH growth models (\eg\ \citealt{Sesana14, Bustamante19, Beckmann24}), and
high-mass systems ($\log[m_{\rm{BH}}] \gtrsim 9$) carry particular diagnostic
power (\citealt{Piotrowska24_hexp}), since most models predict that more
moderate mass systems ($\log[m_{\rm{BH}}] \sim 6-8$) should mostly be rapidly
rotating.  While there are a growing number of systems with both mass and spin
constrained independently, such that the mass--spin plane is becoming
increasingly populated, the majority of these SMBHs lie in the $\log[m_{\rm{BH}}]
\sim 6-8$ range (\citealt{Reynolds21rev}), and the high-mass regime remains
sparsely populated with just two constraints in the $\log[m_{\rm{BH}}] \geq 9$
range (\citealt{Reynolds14, SiskReynes22}).

\begin{figure}
\begin{center}
\hspace*{-0.35cm}
\rotatebox{0}{
{\includegraphics[width=235pt]{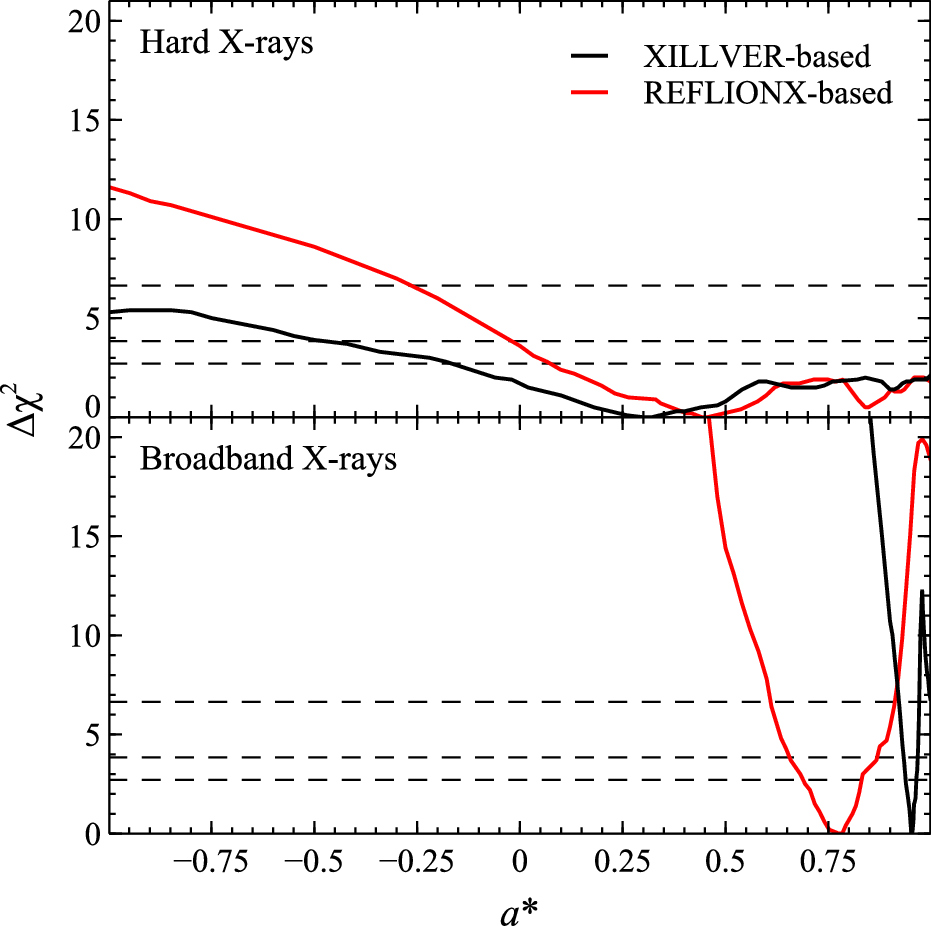}}
}
\end{center}
\vspace*{-0.3cm}
\caption{\textit{Top panel:} the confidence contours for the black hole spin for \pg\
obtained with the best-fit \xillver-based (black) and \reflionx-based reflection models
obtained when considering the hard X-ray data only ($>$2\,keV in the rest-frame of
the source). The horizontal dashed lines represent the $\Delta\chi^{2}$
corresponding to the 90, 95 and 99\% confidence intervals. \textit{Bottom panel:} as
above, but for the best-fit models obtained when fitting the full broadband dataset. In
this case, we show the results for the models that adopt a phenomenological broken
powerlaw emissivity profile, and the \xillver-based model is the variant in which the
\oviii\ flux is allowed to vary.
}
\label{fig_spin}
\end{figure}

For both of the reflection models considered here (\reflionx, \xillver) the best-fit
models for the full broadband \xmm\ + \nustar\ dataset (the models involving a
broken powerlaw emissivity profile, and additionally a variable \oviii\ line strength
in the case of \xillver) both provide constraints on the spin.  However, the best-fit
\xillver-based model prefers a higher spin ($a^* = 0.95 \pm 0.01$ at 90\%
confidence; see Table \ref{tab_broad2}) than the best-fit \reflionx-based model
($a^* = 0.77^{+0.06}_{-0.08}$; see Table \ref{tab_broad1}); the confidence
contours for both models are shown in Figure \ref{fig_spin}. A qualitatively similar
difference can be seen when comparing the lamppost variants as well. As
discussed in \cite{Pottayil24}, this is because the \xillver\ reflection model still
typically has a weaker soft X-ray continuum than \reflionx\ even for the same
density parameter and, as such, the relative contribution of the emission lines to
the overall soft X-ray flux in the reflection spectrum is larger for the \xillver\
model. This means the \xillver\ model requires stronger relativistic blurring than
\reflionx\ in order to produce the smooth soft excess seen in the data, requiring
in turn a higher spin parameter or a steeper emissivity profile (or both).

Despite the above, the 3$\sigma$ confidence intervals ($\Delta \chi^{2} = 9$) for the
best-fit \reflionx-based and \xillver-based models do actually have some degree of
overlap though. Furthermore, the difference in the two spin constraints is of a broadly
similar magnitude to the systematic errors expected to be relevant for spin
measurements ($\Delta a^* \sim 0.1$ for moderate-to-high spins) based on both
parameter recovery simulations relating to spectral modelling (\eg\ \citealt{Bonson16,
Choudhury17, Kammoun18}) and simulations exploring how robust the assumption
that the disc truncates at the ISCO really is (\citealt{Reynolds08}). The level of
disagreement between the two models may therefore not be as severe as it first
seems. Considering the parameter constraints from both models, one may conclude
that reflection models for \pg\ imply the spin of the central supermassive black hole
is at least moderately high ($a^* \gtrsim 0.7$) when the full broadband dataset is
analysed.

It is important to stress, though, that even this combined constraint on the spin
should still be considered model dependent in a more fundamental sense since it is
clearly driven by the assumption that the soft excess is dominated by relativistic
reflection; as discussed above, alternative models are frequently discussed in the
literature (most notably the warm corona model).  The high-energy data
($\gtrsim$2\,keV), where the contribution from relativistic reflection is much less
contentious, do not have sufficient S/N to meaningfully corroborate the broadband
constraints by themselves,  and only provide very weak constraints on the spin,
permitting essentially all positive values of $a^*$ (i.e. prograde spins) for both of
the reflection models considered; confidence contours for the fits that are based
on only the hard X-ray band (Section \ref{sec_hard}) are also shown in Figure
\ref{fig_spin}. Again combining the constraints, the hard X-ray data can only limit
the spin to $a^* \gtrsim -0.15$. The spin constraint for presented by \cite{Mallick25}
-- who treat the soft excess with the warm corona model for \pg\ -- is also similarly
loose,  allowing for the majority of the prograde parameter space ($0.3 \leq a^* \leq
0.93$). As such, the tighter constraints implied by the broadband modelling presented
here may still need to be treated with caution.

\section{Conclusions}
\label{sec_conc}

We have presented the first broadband (\xmm+\nustar) observation of the quasar \pg,
which is powered by the black hole with the largest reverberation-mapped mass
to date: $\log[M_{\rm{BH}} / M_{\odot}]$ = $9.01^{+0.11}_{-0.16}$.  The \xmm\ data,
and in particular the high-resolution RGS spectrum, confirms that this source is
another example of the rare subset of bare AGN. The broadband spectrum shows a
strong soft excess, as well as independent evidence for reflection from the innermost
accretion disc via the presence of a relativistically broadened iron emission line and a
clear Compton hump. 

We use both \xillver-based and \reflionx-based models that allow the density of the
disc to be a free parameter to test whether the soft excess in \pg\ can be explained
with relativistic reflection.  Ultimately we find that both reflection models are able to
do so, but in the case of the \xillver-based models considered we also find that the
strength of the oxygen emission plays a key role in determining whether this is the
case or not. \xillver-based models with the level of oxygen emission included as
standard struggle to fit the broadband data, leaving clear hard excesses in the
residuals (similar to the results found by some previous authors for other bare AGN),
while \xillver-based models that allow for a reduced amount of oxygen emission can
fit the broadband data well. In contrast, the \reflionx-based models can fit the soft
excess without the need for any such modification. Understanding this issue further
will likely be key for determining whether relativistic reflection remains a viable
explanation for the soft excess in general.

The reflection models that do fit the broadband data well imply a high density for
the accretion disc of \logne\ $\sim 18$, though \pg\ is consistent with the loose
anti-correlation between \logne\ and $\log[m_{\rm{BH}} \dot{m}^2]$ seen from
other AGN with reflection-based density constraints. These broadband fits also
provide preliminary constraints on the spin, in combination implying a
moderate-to-high spin of $a^* \gtrsim 0.7$, though here there is formally a bit
more tension between the two reflection models (with our best-fit \reflionx\ model
implying $a^* = 0.77^{+0.06}_{-0.08}$ and our best-fit \xillver\ model implying
$a^* = 0.95 \pm 0.01$).  This result is highly model dependent, though, as it relies
on the soft excess being dominated by relativistic reflection; the hard X-ray data
do not have sufficient S/N to provide a meaningful independent constraint from
the iron line and Compton hump alone (only implying $a^* \gtrsim -0.15$). Further
deep observations will be required to provide a less model-dependent constraint, 
but may be worthwhile given that its high black hole mass places \pg\ in a sparsely
populated part of the spin--mass plane for SMBHs.

\section*{ACKNOWLEDGEMENTS}

The authors would like to thank the reviewer for their positive feedback, which
helped to improve the final version of the manuscript.
DJW acknowledges support from the Science and Technology Facilities Council
(STFC; grant code ST/Y001060/1). 
PK acknowledges support from NASA through the NASA Hubble
Fellowship grant HST-HF2-51534.001-A awarded by the Space Telescope Science
Institute, which is operated by the Association of Universities for Research in
Astronomy, Incorporated, under NASA contract NAS5-26555.
CP is supported by PRIN MUR SEAWIND -- European Union -- NextGenerationEU.
This research has made use of data obtained with \nustar, a project led by Caltech,
funded by NASA and managed by the NASA Jet Propulsion Laboratory (JPL), and has
utilized the \nustardas\ software package, jointly developed by the Space Science
Data Centre (SSDC; Italy) and Caltech (USA).
This research has also made use of data obtained with \xmm, an ESA science mission
with instruments and contributions directly funded by ESA Member States.

%{\it Facilities:} \facility{NuSTAR}, \facility{XMM-Newton}, \facility{Chandra}

\section*{Data Availability}

All of the raw data used in this article are publicly available from ESA's \xmm\ Science
Archive\footnote{https://www.cosmos.esa.int/web/xmm-newton/xsa} and NASA's
HEASARC archive\footnote{https://heasarc.gsfc.nasa.gov/}.

\bibliographystyle{../../mnras}

\bibliography{../../references}

\label{lastpage}

\end{document}